\documentclass[10pt, twocolumn,showpacs,
preprint
]{revtex4}
\usepackage{amsmath,amssymb}
\usepackage{txfonts}
\newcommand{\sn}{\mathrm{sn}}
\newcommand{\cn}{\mathrm{cn}}
\newcommand{\dn}{\mathrm{dn}}
\usepackage{amsmath,amssymb}
\usepackage[dvipdfm]{graphicx}
\usepackage{color}
\usepackage{setspace}

\newtheorem{theorem}{Theorem}

\newtheorem{proposition}[theorem]{Proposition}
\newtheorem{claim}[theorem]{Claim}

\newenvironment{remark}[1][Remark.]{\begin{trivlist}
\item[\hskip \labelsep {\bfseries #1}]}{\end{trivlist}}

\begin{document}
\title{Spectral and formal stability criteria of spatially inhomogeneous 
stationary solutions to the Vlasov equation 
for the Hamiltonian mean-field model}
\author{Shun Ogawa}
\email[E-mail: ]{sogawa@amp.i.kyoto-u.ac.jp}
\affiliation{
	Department of Applied Mathematics and Physics, 
	Graduate School of Informatics, Kyoto University, 
	606-8501 Kyoto, Japan}
\begin{abstract}
Stability of spatially inhomogeneous solutions 
to the Vlasov equation is investigated for the Hamiltonian mean-field model 
to provide the spectral stability criterion and the formal stability criterion 
in the form of necessary and sufficient conditions. 
These criteria determine stability of spatially inhomogeneous solutions 
whose stability has not been decided correctly by using 
a less refined formal stability criterion. 
It is shown that some of such solutions can be found 
in a family of stationary solutions to the Vlasov equation, 
which is parametrized with macroscopic quantities 
and has a two-phase coexistence region in the parameter space. 
\end{abstract}
\maketitle
\section{Introduction}  
\label{sec:intro}

The macroscopic behavior of many-body systems 
depends on whether the interaction is of 
long-range or short-range. 
For many-body systems with short-range interaction, 
the thermodynamic observables such as entropy and 
magnetization are additive and extensive 
but not so for those with long-range interaction. 
The additivity and the extensivity of observables are assumed to hold 
in the equilibrium statistical mechanics and thermodynamics. 
Macroscopic behaviors of many-body systems with 
long-range interaction are quite different from 
those with short-range one 
\cite{AC09, LNP-2002, FB10, JSM-topic}. 
The long-range interaction system is likely to be trapped in quasi-stationary states (QSSs), 
and accordingly a very long time is needed to reach 
the thermal equilibrium state. 
The duration of those QSSs increases according to 
the system size, and diverges if one takes the thermodynamic limit
\cite{JBST, YYY04, JB06, AT11, PdB11, PHC10, PHC12}. 
It is a widely accepted understanding that 
the equilibration is brought about by the finite size effect. 

A way to analyze a Hamiltonian system with long-range interaction 
is to use the Vlasov equation or collisionless Boltzmann equation
\cite{JBST, LPEL}, 
which can be derived by taking the limit 
of $N\rightarrow \infty$, 
where $N$ is the number of elements \cite{BH77, RLD, HS}. 
The QSSs are supposed to be associated with 
stable stationary solutions 
to the Vlasov equation \cite{JBST, AC09}. 
Finding a stability criterion for stationary solutions to the Vlasov equation 
is the first step to investigate QSSs, 
since such a criterion makes it possible to 
decide whether a stationary solution can be a QSS or not.

The stability of solutions to the Vlasov equation has been investigated in
\cite{JBST, Penrose, VAA, DDH85, HEK90, HEK91, SI93, YYY04, JB06, CD09, AC10, ML12}. 
There are several concepts of stability such as the spectral stability, 
the linear stability,  the formal stability, and the nonlinear stability \cite{DDH85}. 
The interest of this paper centers on the spectral stability and the formal stability, 
but the linear stability and the nonlinear stability are not  
touched upon. 
The formal \cite{DDH85, YYY04} and 
spectral \cite{Penrose, CD09, SI93} stability criteria for 
spatially homogeneous solutions have been well known already. 

Meanwhile, the stability of spatially inhomogeneous solutions 
has been investigated in the astrophysics 
\cite{VAA, JBST,HEK90, HEK91} 
since around a half century ago. 
The stability for the spherical galaxy 
is rigorously investigated recently \cite{ML12}.
Antonov's variational principle \cite{JBST, VAA}
particularly gives a necessary and sufficient condition 
for stability of some stationary solution 
by considering stability against  
not all perturbations 
but only accessible perturbations 
called phase preserving perturbations \cite{HEK91}. 
The restriction for the perturbations comes from the fact that 
the Vlasov equation has an infinite number of invariants. 
We note that 
the stability of given stationary state cannot be determined 
by using the stability criterion given in a statement of 
Antonov's variational principle \cite{JBST} practically.

In the context of statistical physics for QSSs, 
the stability of spatially inhomogeneous solutions 
to the Vlasov equation has been studied, say, 
by Campa and Chavanis \cite{AC10}. 
They set up criteria for formal stability both 
in the most refined form and in less refined forms, 
by using the fact that accessible perturbations 
conserve all Casimir invariants at linear order. 
We call the most refined formal stability, 
simply,  
the formal stability in this paper. 
Their formal stability criterion in the most refined form 
requires one to take into account an infinite number of Casimir invariants 
and to detect an infinite number of associated Lagrangian multipliers 
in order to determine the stability of spatially inhomogeneous stationary solutions. 
Their formal stability criterion is hence hard to use. 
In contrast with this, the canonical formal stability criterion 
which is one of the less refined formal stability criteria is of practical use. 
Using the canonical formal stability criterion, 
one can check stability of a stationary state against a perturbation which keeps 
the normalization condition but may break the energy conservation 
and other Casimir invariant conditions. 
Though the criterion for canonical formal stability is stated as 
a necessary and sufficient condition, 
it is just a sufficient condition for the formal stability.  
It is to be expected that a criterion for the formal stability is found out 
in the form of necessary and sufficient condition without 
reference to an infinite number of quantities such as 
Lagrangian multipliers.

This article deals with the Hamiltonian mean-field (HMF) model \cite{JM82, MA} 
with the anticipation stated above. 
The HMF model is a simple toy model which shows typical long-range features. 
For instance, the HMF model has been used for investigating 
the nonequilibrium phase transitions \cite{PHC06, AA07, FS11, SO11}, 
the core-halo structure \cite{RP11}, 
the creation of small traveling clusters \cite{JB09}, 
the construction of traveling clusters \cite{YYY11},  
and a relaxation process with long-range interactions 
\cite{YYY04, JB06}. 
Moreover, the HMF model allows one to perform 
theoretical study on dynamics near spatially inhomogeneous stationary solutions 
to the Vlasov equation 
by the use of the dispersion function which can be 
explicitly written out for the HMF model. 
For instance, the dynamics of a perturbation around the spatially inhomogeneous 
stationary solution \cite{BOY10} and the algebraic damping to a QSS \cite{BOY11} have been 
investigated theoretically and numerically by using the HMF model. 
Further, the linear response to the external field is studied 
in an explicit form \cite{SO12} for a spatially inhomogeneous QSS. 
In those studies, the stability of the spatially inhomogeneous solutions 
have been assumed to hold, and then it is worthwhile to give 
an explicit form of necessary and sufficient condition for 
the stability of the spatially inhomogeneous stationary solutions. 
The aim of this article is to find spectral and 
formal stability criteria 
for spatially inhomogeneous stationary solutions.  
The spectral stability criterion is derived  
by means of the dispersion relation. 
The formal stability criterion is obtained by using 
the same idea as exhibited in \cite{AC10}.  
The criterion we are to find by using the angle-action variables 
is free from an infinite number of Lagrangian multipliers, 
and is stated in the form of a necessary and sufficient condition, 
which allows us to look into the stability of spatially inhomogeneous solutions 
in an accessible manner. 

This article is organized as follows.
Section \ref{sec:stability} contains 
a brief review of the two kinds of stabilities of 
a fixed point of a dynamical system. 
The nonlinear and linearized Vlasov equations for the HMF model are 
introduced in Sec. \ref{sec:LVE}. 
The spectral stability criterion for spatially homogeneous solutions to the 
Vlasov equation is given in Sec. \ref{sec:hom} 
in a rather simple method than that already known. 
By using the same method, the spectral stability criterion 
for spatially inhomogeneous solutions is obtained 
in Sec. \ref{subsec:inhom-spec}. 
The formal stability criterion 
for spatially inhomogeneous solutions is 
derived in Sec. \ref{subsec:inhom-formal}. 
In Sec. \ref{subsec:water-bag}, 
we look into 
stability of a spatially inhomogeneous 
water-bag distribution by using the obtained criterion. 
Section \ref{sec:example} gives 
an example which shows that the present stability criterion 
is of great use. 
It is shown that there is a family of stationary solutions 
whose stability cannot be judged correctly by using the canonical formal stability criterion 
but can be done by the criterion given in this article. 
Section \ref{sec:conclusion} is devoted to a summary and a discussion for generalization.

\section{Spectral stability and formal Stability}
\label{sec:stability}
We start with a brief review of 
definitions of spectral stability and formal stability, 
following Holm et al. \cite{DDH85}. 
Let $X$ be a normed space. 
Suppose that a dynamical system 
is given by the equation, 
\begin{equation}
	\label{eq:dyn-sys}
	\frac{dx}{dt} = f(x),\quad x \in X. 
\end{equation}
Let $x_\ast$ be a fixed point of this system, $f(x_\ast) = 0$. 
Then, the linearized equation around $x_\ast$ is 
expressed as 
\begin{equation}
	\label{eq:lin-sys}
	\frac{d \xi}{dt} = Df(x_\ast) [\xi], 
\end{equation}
where $Df(x_\ast)$ is a linear operator 
derived from $f$ at $x_\ast$. 
The spectral stability and the formal stability of 
the fixed point $x_\ast$ 
are defined as follows:
\begin{itemize}
	\item  
	The fixed point $x_\ast$ is said to be {\it spectrally stable}, 
	if the linear operator $Df(x_\ast)$ has no spectrum 
	with positive real part. 
	In addition, if the linear operator $Df(x_\ast)$ 
	has an eigenvalue with vanishing real part, 
	$x_\ast$ is called {\it neutrally spectrally stable}. 
	The fixed point $x_\ast$ is said to be spectrally unstable 
	when there exists a spectrum with positive real part. 
		
	\item The fixed point $x_\ast$ is said to be {\it formally stable}, 
	if a conserved functional $\mathcal{F}[x]$ takes 
	a critical value at $x = x_\ast$ and further 
	the second variation of $\mathcal{F}$ at $x_\ast$ is 
	negative (or positive) definite. 
	The fixed point $x_\ast$ is said to be neutrally formally stable 
	if the second variation of $\mathcal{F}$ at $x_\ast$ is 
	negative (resp. positive) semi-definite but not 
	negative (resp. positive) definite.  
	Further, the fixed point $x_\ast$ said to be 
	formally unstable if the second variation of $\mathcal{F}$ at $x_\ast$ is 
	not negative (or positive) semi-definite. 
\end{itemize}
We note that the formal stability can be defined for $x^\ast$ 
which is a critical point of $\mathcal{F}$ under some constraints 
coming from invariants of the dynamical system in question. 

If the dynamical system in question is infinite dimensional, 
the fixed point $x_\ast$ is occasionally called a stationary state.
We note that the definition of neutral spectral stability is different from 
the original one in \cite{DDH85}. The detail of our footing for 
stability analysis is exhibited in 
Appendix \ref{sec:neutral}. 
According to \cite{DDH85}, the neutrally spectrally stable solution 
is spectrally stable, but 
the neutrally formally stable solution is not formally stable. 

\section{Vlasov equation for Hamiltonian mean-field model}
\label{sec:LVE}
The Hamiltonian mean-field (HMF) model \cite{MA, JM82} 
for $N$ unit mass particles on the unit circle $S^1$ 
has the Hamiltonian given by
\begin{equation}
	\label{eq:HMF}
	\begin{split}
		&H_N = \sum_{i=1}^N \frac{p_i^2}{2} 
		+ \frac{1}{2N}\sum_{i,j=1}^N\left(1 - \cos(q_i - q_j)\right),\\
		&p_i \in \mathbb{R}, \quad 
		q_i \in [-\pi,\pi), \quad i=1,2,\cdots N.  
	\end{split}
\end{equation}
In the limit of $N$ tending to infinity, 
the time evolution of the HMF model 
can be described in terms of a single-body distribution $f$ 
on the $\mu$-space which coincides with $S^1 \times \mathbb{R}$. 
The single-body distribution $f$ is known to evolve according to 
the Vlasov equation, 
\begin{equation}
	\label{eq:vla}
	\frac{\partial f}{\partial t} + \{\mathcal{H}[f], f \} = 0, 
\end{equation}
where $\mathcal{H}[f]$ is the effective single-body Hamiltonian defined to be 
\begin{equation}
\label{eq:vla2}
	\begin{split}
	&\mathcal{H}[f] = \frac{p^2}{2} + \mathcal{V}[f](q,t) , \\
	&\mathcal{V}[f]=
	-\int_{-\pi}^{\pi} dq' \cos(q-q')\int_{-\infty}^{\infty} f(q',p',t)~dp',\\ 
	&p \in \mathbb{R},\quad q \in [-\pi, \pi), 
	\end{split}
\end{equation}
and where $\{a,b\}$ is the Poisson bracket given by 
\begin{equation}
	\label{eq:PB}
	\{a,b\} = \frac{\partial a}{\partial p}\frac{\partial b}{\partial q}
	-\frac{\partial a}{\partial q}\frac{\partial b}{\partial p}.  
\end{equation} 
By averaging the distribution $f$ in the $x$- and $y$-directions, 
the order parameter $\vec{M}[f] = \left(M_x[f], M_y[f]\right)^T$ is defined to be 
\begin{equation}
	\begin{split}
		M_x[f](t) &= \iint_\mu \cos q~f(q,p,t)~dq dp,\\
		M_y[f](t) &= \iint_\mu \sin q~f(q,p,t)~dq dp, 
	\end{split}
\end{equation}
where the symbol $\mu$ denotes the whole $\mu$-space, 
$S^1 \times \mathbb{R}$. 
In the Vlasov dynamics, a functional  
\begin{equation}
	\label{eq:en-cas0}
	\mathcal{Q}[f]\equiv \iint_\mu Q\left(f(q, p, t)\right) ~dq dp
\end{equation}
is conserved for any function $Q$, 
and such a functional is called a Casimir invariant.

Let $f_0$ denote a stationary solution to the Vlasov equation. 
Then a small perturbation $f_1$ around $f_0$ is shown to 
obey, in some timescale, 
the linearized Vlasov equation 
\begin{equation}
	\label{eq:lin}
	\begin{split}
		\frac{\partial f_1}{\partial t} =& \hat{\mathcal{L}} f_1, \\
		\hat{\mathcal{L}}f_1 :=
		&- \{\mathcal{H}[f_0], f_1\} - \{\mathcal{V}[f_1] , f_0\}. 
	\end{split}
\end{equation}  
This equation can be analyzed by means of the 
Fourier-Laplace transformation. 
For the sake of physical interpretation, 
we define the Laplace transform of a function $g(t)$ to be
\begin{equation}
	\tilde{g}(\omega)
	= \int_0^\infty g(t)e^{i\omega t} dt, \quad 
	{\rm Im} \omega > 0.
\end{equation}
Through the Fourier series expansion with respect to $q$ 
and the Laplace transformation with respect to $t$, 
Eq. \eqref{eq:lin} is brought into 
the dispersion relation $D(\omega) = 0$ \cite{LDL, LPEL}. 
The explicit form of the dispersion relations 
for spatially homogeneous stationary states and 
for spatially inhomogeneous stationary states 
will be exhibited in Sec. \ref{sec:hom} 
and in Sec. \ref{sec:inhom}, respectively. 
We call $D(\omega)$ a dispersion function, 
which is said to be a dielectric function in the context of the plasmas physics. 
A root $\omega$ of this dispersion relation with positive 
imaginary part is 
in one-to-one correspondence 
with the eigenvalue $-i\omega$ of 
the linear operator $\hat{\mathcal{L}}$.  
The detail of this fact is reviewed in Appendix \ref{sec:app1}. 
It then follows that the stationary solution $f_0$ is 
spectrally unstable 
if the dispersion relation $D(\omega) = 0$ has a root in the 
upper half $\omega$-plane. 

Though the domain of the dispersion function 
$D(\omega)$ is the upper half $\omega$-plane, 
it can be analytically continued 
to the lower half $\omega$-plane \cite{LDL, LPEL}. 
The root of the dispersion relation on the lower half plane 
causes the exponential damping of the order parameter for the small perturbation 
$\vec{M}[f_1] = \left(M_x[f_1], M_y[f_1] \right)^T$, 
which is called the Landau damping \cite{LDL, LPEL, SHS92}. 
Such a root is not an eigenvalue, but is called 
a resonance pole, a Landau pole, or a fake eigenvalue. 

We remark on embedded eigenvalues of 
the linear operator $\hat{\mathcal{L}}$ on the imaginary axis. 
The linear operator $\hat{\mathcal{L}}$ may have 
continuous spectra lying on the imaginary axis \cite{JDC89}. 
An ``eigenvalue'' with zero real part is occasionally 
embedded in continuous spectra and 
such an ``eigenvalue'' is called an embedded eigenvalue.

Let a stationary solution $f_0$ be spatially inhomogenous, i.e., 
$\left| \vec{M}[f_0] \right|\neq 0$. 
Owing to the rotational symmetry of the HMF model, 
there must be a mode $f_1^0$ associated with the zero embedded eigenvalue 
of the linearized Vlasov operator. 
The direction of a perturbation $\vec{M}[f_1^0]$ 
is perpendicular to the direction of the order parameter $\vec{M}[f_0]$. 
This fact is consistent with the fact that the order parameter changes its direction, 
keeping its radius, in the presence of an infinitesimal external field which 
is perpendicular to the order parameter \cite{SO12}.  
When the nonequilibrium phase transition \cite{PHC06, AA07, FS11, SO11} is of interest, 
the phases are defined by the modulus of the order parameter, 
and hence the direction of the order parameter is not questioned. 
At present, we mainly focus on the stability against the 
perturbation parallel to the order parameter $\vec{M}[f_0]$.

\section{Stability criterion 
for spatially homogeneous stationary solution}
\label{sec:hom}
As long as the linear operator $\hat{\mathcal{L}}$ 
defined in \eqref{eq:lin} is concerned, the spectral stability condition 
for a stationary solution $f_0$ to the Vlasov equation can be compactly stated; 
if there is no eigenvalue of $\hat{\mathcal{L}}$, 
the stationary solution $f_0$ is said to be spectrally stable. 
Since the spectrum of $\hat{\mathcal{L}}$ 
consist of eigenvalues on $\mathbb{C}\setminus i\mathbb{R}$, 
continuous spectra on the imaginary axis, 
and the embedded eigenvalue on the imaginary axis \cite{JDC89}, 
only the eigenvalues are able to contribute to the spectral instability. 
On account of this fact, the spectral stability criterion is stated as follows: 
\begin{proposition}
\label{prop1}
Let $f_0(p)$ be a spatially homogeneous stationary solution to the Vlasov equation, 
which is assumed to be 
smooth, even, and unimodal, and 
further the derivative $f_0'(p)$ of which 
is assumed to have the support $\mathbb{R}$. 
Then the  stationary solution $f_0$ is spectrally stable, 
if and only if $f_0$ satisfies the inequality 
\begin{equation}
	\label{eq:hom-crit}
	\mathcal{I}[f_0] = 1 + \pi \int_{-\infty}^{\infty} \frac{f_0'(p)}{p}~dp \geq 0. 
\end{equation}
\end{proposition}

We note that $f_0'(p)/p$ has no singurality for all $p \in \mathbb{R}$ 
on account of the assumption that $f_0$ is smooth and even. 
Though the inequality \eqref{eq:hom-crit} can be derived 
by using the Nyquist's method \cite{CD09, SI93, Penrose}, 
we introduce a method other than the Nyquist's method 
to prove this proposition. 

For a spatially homogeneous stationary solution $f_0(p)$, 
the dispersion relation $D(\omega) = 0$ 
with ${\rm Im} \omega > 0$ is put in the form \cite{AC09}
\begin{equation}
	\label{eq:disp-hom}
	D(\omega) = 
	1 + \pi \int_{-\infty}^{\infty} \frac{f_0'(p)}{p - \omega}~dp = 0, 
	\quad {\rm Im} \omega>0. 
\end{equation}
The dispersion function is continued to $\omega = 0$ 
from the upper half $\omega$-plane, 
by taking the limit $D(0) = \lim_{\epsilon \to 0+} D(i\epsilon)$. 
Noting that $f_0'(p)/ p$ has no singularity, 
we obtain $D(0) = \mathcal{I}[f_0]$, 
since the integrand in 
\eqref{eq:disp-hom} has no singularity when 
$\omega = 0$. 
We put $\omega \in \mathbb{C}$ in the form 
$\omega = \omega_{\rm r} + i \omega_{\rm i}$ with 
$\omega_{\rm r} \in \mathbb{R}$ and $\omega_{\rm i} > 0$. 
When the dispersion relation \eqref{eq:disp-hom} 
is satisfied by some $\omega$ with ${\rm Im} \omega > 0$, the imaginary part of $D(\omega)$ is zero, 
so that one has
\begin{equation}
	\label{eq:hom-disp-im}
	\begin{split}
		\mathrm{Im} D(\omega) 
		&= 
		\pi \omega_{\rm i} \int_{-\infty}^{\infty} 
		\frac{f_0'(p)}{(p-\omega_{\rm r})^2 + \omega_{\rm i}^2} dp\\
		&= 
		4\pi \omega_{\rm i} \omega_{\rm r}
		\int_0^{\infty} 
		\frac{pf_0'(p)~dp}{\left((p-\omega_{\rm r})^2 + \omega_{\rm i}^2 \right)\left((p+\omega_{\rm r})^2 + \omega_{\rm i}^2 \right)} \\
		&= 0.
	\end{split}
\end{equation}
Since $f_0(p)$ is an even unimodal function, 
$p f_0'(p) < 0$ for all $p> 0$. 
The integral in \eqref{eq:hom-disp-im} is to be negative value, 
and \eqref{eq:hom-disp-im} implies that $\omega_{\rm r} = 0$, 
since $\omega_{\rm i} > 0$.  
Conversely, if $\omega_{\rm r} = 0$, 
the equality ${\rm Im} D(\omega) = 0$ holds true. 
The condition $\omega_{\rm r} = 0$ is then equivalent to 
the condition $\eqref{eq:hom-disp-im}$.

Now, on account of the fact 
$p f_0'(p)$ is negative for all $p \neq 0$, 
and the dispersion function satisfies the inequality
\begin{equation}
	\label{eq:hom-disp-re}
	D(i\omega_{\rm i}) = 
	1 + \pi \int_{-\infty}^{\infty} 
	\frac{p f_0'(p)}{p^2 + \omega_{\rm i}^2} dp
	\geq \mathcal{I}[f_0],
\end{equation}
for all $\omega_{\rm i} \geq 0$, where the equality is satisfied if and only if $\omega_{\rm i} = 0$. 
This is because $D(i \omega_{\rm i})$ becomes $\mathcal{I}[f_0]$ for $\omega_{\rm i} = 0$ 
and because $D(i\omega_{\rm i})$ is a continuous and 
strictly increasing function with respect to $\omega_{\rm i}$. 
It then follows that 
if $D(\omega) = 0$ with $\omega = i \omega_{\rm i}$, or equivalently, 
if $\hat{\mathcal{L}}$ has an eigenvalue
$-i \omega$ with ${\rm Im} \omega > 0$, then 
$\mathcal{I}[f_0] < 0$. 
Conversely, if the inequality $\mathcal{I}[f_0] < 0$ is satisfied, 
there exists a positive $\omega_{\rm i}$ 
such that $D(i\omega_{\rm i}) = 0$. 
In fact, $D(i \omega_{\rm i})$ is strictly increasing in $\omega_{\rm i}$ 
with $D(0) = \mathcal{I}[f_0]$ 
and $D(i \omega_{\rm i}) \rightarrow 1$ as $\omega_i \rightarrow \infty$. 
This means that $\hat{\mathcal{L}}$ has an unstable eigenvalue 
if and only if 
$\mathcal{I}[f_0] < 0$.
Thus we have proved the spectral stability criterion \eqref{eq:hom-crit}.

In comparison with the spectral stability criterion \eqref{eq:hom-crit}, 
the formal stability criterion \cite{YYY04} is given by
\begin{equation}
	\label{eq:hom-for}
	\mathcal{I}[f_0] > 0.
\end{equation}
This inequality means that 
$f_0$ is spectrally stable but not neutrally spectrally stable. 
This is because if $D(0) =\mathcal{I}[f_0]= 0$, 
the linear operator $\hat{\mathcal{L}}$ 
has an embedded eigenvalue $0$, 
and hence $f_0$ is neutrally spectrally stable. 

\section{Stability criteria for  
spatially inhomogeneous stationary solution}
\label{sec:inhom}
In this section, we will give 
necessary and sufficient conditions 
for the spectral stability and for 
the most refined formal stability of spatially inhomogeneous 
stationary solutions to the Vlasov equation. 
We call the most refined formal stability, simply, 
the formal stability as we have already mentioned 
in the introduction. 

A spectral stability criterion 
for spatially inhomogeneous solutions 
can be given in an explicit form 
by performing the same procedure as that adopted in the last section. 

Furthermore, the formal stability criterion can be worked out 
if all the Casimir invariants are taken into account. 
For spatially inhomogeneous stationary solutions, 
Campa and Chavanis  \cite{AC10} have given 
the formal stability criterion. 
However, no one has these criteria explicitly, 
since one needs to detect values of 
an infinite number of Lagrangian multipliers. 
We can avoid a puzzle to detect an infinite number of 
Lagrangian multipliers if we use the angle-action coordinates 
in stability analysis. 

We denote the single-body energy by
\begin{equation}
\label{e:1bodyE}
	\mathcal{E}(q,p) = p^2/2 - M_0 \cos q, 
\end{equation}
where on account of the rotational symmetry of the HMF model, 
the order parameter has been set 
$\vec{M}_0 = (M_0, 0)$ with 
\begin{equation}
	M_0 = \iint_\mu \cos q~f_0(q,p)~dq dp. 
\end{equation}

\subsection{Angle-action coordinates for HMF model}

Before analyzing the stability of spatially inhomogeneous stationary solutions, 
we review the angle-action coordinates $(\theta, J)$ 
for the HMF model. 
The detail of constructing the angle-action coordinates 
can be found in \cite{BOY10}.

\begin{figure}[t]
	\begin{center}
		\includegraphics[width=8.5cm]{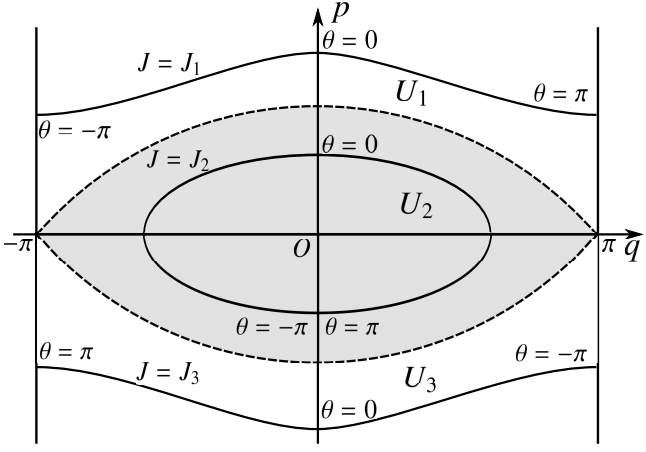}
		\caption{
		We illustrate the angle-action variables in regions 
		$U_i$ ($i=1, 2, 3)$ in the $\mu$ space. 
		The broken curve is a separatrix.  
		The region $U_2$ is the gray region surrounded by the septaratix.
		The solid curves in each regions are tragectries of 
		the dynamics induced by the effective single-body Hamiltonian 
		$\mathcal{H}[f_0](q,p) = p^2/2 - M_0 \cos q$ .
		}
		\label{fig:action-angle}
	\end{center}
\end{figure}

To construct a bijective mapping of 
$(q,p)$ to $(\theta, J)$,  
we divide the $\mu$-space into three regions, 
$U_1, U_2$, and $U_3$, which are defined, respectively, as 
\begin{equation}
	\begin{split}
		U_1 &= \left\{(q,p)~|~\mathcal{E}(q,p) > M_0,p>0\right\},\\
		U_2 &= \left\{(q,p)~|~|\mathcal{E}(q,p)| < M_0\right\},\\
		U_3 &= \left\{(q,p)~|~\mathcal{E}(q,p) > M_0,p<0\right\}.
	\end{split}
\end{equation}
According to this division of the $\mu$-space, 
we prepare the sets $V_1, V_2$, and $V_3$ defined to be 
\begin{equation}
	\begin{split}
		V_1 &= \left\{(\theta_1,J_1)~|~\theta_1\in [-\pi,\pi), J_1>4\sqrt{M_0}/\pi\right\},\\
		V_2 &= \left\{(\theta_2,J_2)~|~\theta_2\in [-\pi,\pi), 0<J_2<8\sqrt{M_0}/\pi\right\},\\
		V_3 &= \left\{(\theta_3,J_3)~|~\theta_3\in [-\pi,\pi), J_3>4\sqrt{M_0}/\pi\right\},
	\end{split}
\end{equation}
respectively. 
Then, the maps $(q,p)\mapsto (\theta_i, J_i): U_i \rightarrow V_i$, for $i = 1, 2, 3, $ 
are bijective. 
We illustrate the angle-action variables in three regions 
$U_1$, $U_2$, and $U_3$ in Fig. \ref{fig:action-angle}. 
Since we are interested in integration over the $\mu$-space, 
we do not have to mention more on the boundaries of $U_i$. 

According to these bijections, a function $g$ 
whose arguments are the angle-action variables $(\theta, J)$ 
is denoted by
\begin{equation}
	\label{eq:g}
	g(\theta,J) =
	\begin{cases}
		g_1(\theta_1, J_1),\quad (\theta, J) \in V_1,\\
		g_2(\theta_2, J_2),\quad (\theta, J) \in V_2,\\
		g_3(\theta_3, J_3),\quad (\theta, J) \in V_3,
	\end{cases}
\end{equation}
respectively. 
We will omit the subscript $i$ if no confusion arises. 
For notational simplicity, we denote the integral of the function \eqref{eq:g} 
over the whole $\mu$-space by the left-hand side of the following equation
\begin{equation}
\label{eq:intg}
	\iint_\mu g(\theta, J)~ d\theta dJ \equiv 
	\sum_{i=1,2,3}\iint_{V_i} g_i(\theta_i, J_i)~d\theta_i dJ_i . 
\end{equation}
In a similar manner, the integration of a function $f(J)$ is put in the form, 
\begin{equation}
	\label{eq:intf}
	\begin{split}
		\int_L f(J)~dJ &\equiv
		\int_{4\sqrt{M_0}/\pi}^\infty f_1(J_1)~dJ_1\\ 
		+&\int_0^{8\sqrt{M_0}/\pi} f_2(J_2)~dJ_2 
		+\int_{4\sqrt{M_0}/\pi}^\infty f_3(J_3)~dJ_3.
	\end{split}
\end{equation}
In the later part of this article, 
the monotonicity of a function $f(J)$ with respect to $J$ means 
the monotonicity of functions $f_i(J_i)$ with respect to $J_i$ 
for each $i =1, 2, 3, $ respectively.

\subsection{Spectral stability criterion}
\label{subsec:inhom-spec}
We derive a necessary and sufficient condition for 
a spatially inhomogeneous stationary solution $f_0$ 
to the Vlasov equation to be spectrally stable, which is stated as follows: 
\begin{proposition}
\label{prop2}
Let $f_0$ be a spatially inhomogeneous stationary solution to the Vlasov equation, 
which is assumed to depend on the action $J$ 
only through the single-body energy $\mathcal{E}(J)$ in such a manner that  
\begin{equation}
	\label{eq:stationary-solution}
	f_0 (q,p)= \tilde{f}_0\left(J(q,p)\right) = \hat{f}_0\left(\mathcal{E}(q,p)\right). 
\end{equation}
Further, $\tilde{f}_0\left(J\right)$ and 
$\hat{f}_0\left(\mathcal{E}\right)$ are assumed to be 
strictly decreasing with respect to $J$ and $\mathcal{E}$, respectively. 
A further assumption is that $d\hat{f}_0\left(\mathcal{E}\right)/d\mathcal{E}$ is 
continuous with respect to $\mathcal{E}$. 
Such a stationary solution $f_0(q,p)$ is spectrally stable, if and only if
\begin{equation}
	\label{eq:inhom-crit}
	\begin{split}
	\mathcal{I}[f_0] 
	 =&
	1 +  \int_{-\pi}^{\pi} dq \cos^2 q
	\int_{-\infty}^{\infty} \frac{1}{p}\frac{\partial f_0}{\partial p}(q,p)~dp \\
	&- 2\pi \int_L \frac{{\tilde{f}_0}'(J)}{\Omega(J)} |C^0(J)|^2 dJ\geq 0, 
	\end{split}
\end{equation}
where $(\theta,J)$ are the angle-action coordinates and 
$\Omega(J) \equiv d\mathcal{E}(J)/dJ$, 
and where $C^n(J)$ is defined by
\begin{equation}
	\label{eq:cn}
	C^n(J) \equiv 
	\frac{1}{2\pi}\int_{-\pi}^{\pi} \cos q(\theta, J) e^{-in \theta} d\theta,
	\quad n \in \mathbb{Z}. 
\end{equation}
\end{proposition}

Before proving Prop. \ref{prop2}, 
we note that all distributions 
such as \eqref{eq:stationary-solution} 
are stationary solutions to the Vlasov equation. 
The monotonicity of $f_0$ 
in $J$ is satisfied for stationary solutions 
which are obtained as solutions to 
a variational equation associated with
an optimization problem 
such as the maximization of the entropy or 
the minimization of the free energy. 
Then the assumption imposed on 
$f_0$ in Prop. \ref{prop2} is not too restrictive, 
and has some physical relevance. 
We note also that 
$\tilde{f}_0'(J)/ \Omega(J)$ is finite for all $J$ since 
$d\hat{f}_0\left(\mathcal{E}\right)/d\mathcal{E}$ has no singularity.

The proof of Prop. \ref{prop2} can be performed in a similar manner to 
that applied to Prop. \ref{prop1}, though the $f_0$ is 
spatially inhomogeneous 
in the present proof. 
We divide the stability analysis into two, 
one of which deals with stability against the perturbation 
in the direction parallel to the order parameter $\vec{M}_0 = (M_0, 0)^T$ 
and the other with stability against the perturbation in the direction perpendicular to $\vec{M}_0$.

We first analyze the stability against the perturbation 
in the direction parallel to the 
order parameter $\vec{M}_0 = \left(M_0, 0\right)^T$. 
The dispersion function in this case is put in the form 
\begin{equation}
	\label{eq:inhom-disp}
	D_x(\omega) = 1 + 2\pi \sum_{m \in \mathbb{Z}} \int_L 
	\frac{m {\tilde{f}_0}'(J)}{m\Omega(J) - \omega}|C^m(J)|^2 dJ, 
	\quad {\rm Im \omega > 0}, 
\end{equation}
and the dispersion relation is given by $D_x(\omega) = 0$ \cite{BOY10}. 
When ${\rm Im} \omega > 0$, the term of $m = 0$ in \eqref{eq:inhom-disp} 
vanishes and 
Eq. \eqref{eq:inhom-disp} is arranged as 
\begin{equation}
	\label{eq:inhom-disp2}
	D_x(\omega) = 1 + 2\pi \sum_{m \in \mathbb{Z}\setminus \{0\}} 
	\int_L \frac{{\tilde{f}_0}'(J)}{\Omega(J) - \omega/m}|C^m(J)|^2 dJ, 
	\quad {\rm Im \omega > 0}. 
\end{equation}
We here note that $D_x(0)$ is defined as 
$D_x(0) = \lim_{\epsilon \to 0+} D_x(i\epsilon)$, and 
\begin{equation}
	\label{eq:aaD0}
	D_x(0) 
	=1 + 2\pi \sum_{m\neq 0} 
	\int_L \dfrac{\tilde{f}_0'(J)}{\Omega(J)} |C^m(J)|^2 dJ, 
\end{equation}
since the integrand in it has no singularity.

If there exists $\omega$ such that
$D_x(\omega) = 0$ with ${\rm Im} \omega > 0$, then one has 
${\rm Im} D_x(\omega) = 0$, 
which is written out as 
\begin{widetext}
\begin{equation}
	\label{inhom-disp-im}
	\begin{split}
		\mathrm{Im}D_x(\omega) &= 
		\sum_{m \in \mathbb{Z} \setminus\{0\}} 
		\frac{2\pi}{m} \omega_{\rm i} 
		\int_{L} \frac{|C^m(J)|^2 \Omega(J) \tilde{f}_0'(J)}
		{(\Omega(J) - \omega_{\rm r}/m)^2 + (\omega_{\rm i}/m)^2}dJ
		\\
		&= 
		\sum_{m \in \mathbb{N}} \left[ 
		\frac{2\pi}{m} \omega_{\rm i} 
		\int_{L} \frac{|C^m(J)|^2 \Omega(J) \tilde{f}_0'(J)}
		{(\Omega(J) - \omega_{\rm r}/m)^2 + (\omega_{\rm i}/m)^2}dJ
		-\frac{2\pi}{m} \omega_{\rm i} 
		\int_{L} \frac{|C^{-m}(J)|^2 \Omega(J) \tilde{f}_0'(J)}
		{(\Omega(J) + \omega_{\rm r}/m)^2 + (\omega_{\rm i}/m)^2}dJ
		\right]
		\\
		&=
		\sum_{m\in \mathbb{N}} 
		\frac{8\pi}{m} \omega_{\rm r} \omega_{\rm i} 
		\int_L \frac{|C^m(J)|^2 \Omega(J) \tilde{f}_0'(J)}
		{\left\{(\Omega(J) - \omega_{\rm r}/m)^2 + (\omega_{\rm i}/m)^2\right\}
		\left\{(\Omega(J) + \omega_{\rm r}/m)^2 + (\omega_{\rm i}/m)^2\right\}}dJ 
		= 0, 
	\end{split}
\end{equation}
\end{widetext}
where we have used the fact $C^m(J) = C^{-m}(J)^\ast$ which is derived from \eqref{eq:cn}. 
Since $|C^m(J)|^2 > 0$ for some $m \in \mathbb{N}$, and $\Omega(J) > 0$ and 
$\tilde{f}_0'(J) < 0$ for all $J$, 
the integrals in \eqref{inhom-disp-im} give negative values. 
Then \eqref{inhom-disp-im} yields $\omega_{\rm r} = 0$, since $\omega_{\rm i} > 0$. 
We thus have shown that if 
$\omega = \omega_{\rm r} + i \omega_{\rm i}$ with $\omega_{\rm i} > 0$ 
is a root of $D_x(\omega) = 0$ then $\omega_{\rm r} = 0$. 

On account of $\omega_{\rm r} = 0$, the dispersion relation reduces to 
\begin{equation}
	\label{inhom-disp-re}
	\begin{split}
	D_x(i \omega_{\rm i}) &= 
	1 + 4\pi\sum_{m\in \mathbb{N}} \int_L \frac{\Omega(J)\tilde{f}_0'(J)}
	{\Omega(J)^2 + (\omega_{\rm i}/m)^2}
	 |C^m(J)|^2 dJ\\
	 &= 0. 
	 \end{split}
\end{equation}
The function $D_x(i \omega_{\rm i})$ is a strictly increasing continuous function 
of $\omega_{\rm i}$, and converges to 1, $D_x(i \omega_{\rm i}) \rightarrow 1$, 
as $\omega_{\rm i} \rightarrow \infty$. 
This implies that if $D_x(0) < 0$, 
there is a positive number $\omega_{\rm i} > 0$ such that $D_x(i\omega_{\rm i}) = 0$. 
Put another way, if $D_x(0) < 0$, there is an $\omega$ such that $D_x(\omega) = 0$, 
${\rm Im} \omega > 0$. 
The converse is also shown by taking the contraposition of that 
there is no root $\omega$ of $D(\omega)$ with ${\rm Im} \omega > 0$ 
if $D_x(0) \geq 0$. 
We hence conclude that 
there is no unstable eigenvalue for the perturbation whose direction is 
parallel to the order parameter $\vec{M}_0 = (M_0, 0)^T$, 
if and only if $D_x(0) \geq 0$.
If $D_x(0) = 0$, the operator $\hat{\mathcal{L}}$ 
has an embedded eigenvalue $0$, so that $f_0$ is neutrally spectrally stable. 

To derive the spectral stability criterion \eqref{eq:inhom-crit}, 
we have only to prove the relation, $D_x(0) = \mathcal{I}[f_0]$, 
which can be done by performing 
the same procedure as that 
carried out in Appendix C of  \cite{SO12}. 
According to Appendix B of \cite{BOY10},
the function $\cos q$ is expressed as
\begin{equation}
	\label{eq:cosk}
	\cos q\left(\theta, J(k)\right) 
	= 
	\begin{cases}
		\vspace{5mm}
		\displaystyle
		1 - 2k^2 \sn^2\left(\frac{2K(k)}{\pi} \theta, k\right), &\quad k< 1, \\
		\displaystyle
		1- 2 \sn^2 \left(\frac{K(1/k)}{\pi} \theta, \frac{1}{k} \right),&\quad k>1,
	\end{cases}
\end{equation}
where $K(k)$ is the complete elliptic integral of the first kind \cite{MAIS72}, 
and where $k$ is defined as 
\begin{equation}
	\label{eq:k-def}
	k \equiv \sqrt{\frac{\mathcal{E}(J) + M_0}{2M_0}}.
\end{equation}
Owing to the periodicity of the Jacobian elliptic functions \cite{MAIS72}, 
the function $\cos q(\theta, J)$ 
is $2\pi$-periodic with respect to $\theta$.  
Then, from Parseval's equality, we obtain
\begin{equation}
    \label{eq:Parseval}
    \begin{split}
        2\pi \sum_{m \neq 0} |C^m(J)|^2 
        &= \int_{-\pi}^\pi \cos^2 q(\theta, J) ~d\theta 
        - 2\pi |C^0(J)|^2.
    \end{split}
\end{equation}
By using this equation, 
we rewrite the second terms in the 
right-hand side of \eqref{eq:aaD0} as
\begin{equation}
    \label{eq:intDx0}
	\int_L dJ \dfrac{\tilde{f}_0'(J)}{\Omega(J)} \int_{-\pi}^{\pi} \cos^2 q(\theta, J) ~d\theta 
	- 2 \pi \int_L \dfrac{\tilde{f}_0'(J)}{\Omega(J)} |C^0(J)|^2~dJ.
  \end{equation}
Keeping in mind the fact that the stationary solution $f_0$ 
to the Vlasov equation depends on the action $J$ only through 
a single-body energy $\mathcal{E}$, 
we arrange the first term of \eqref{eq:intDx0} as 
\begin{equation}
	\label{eq:int1Dx0}
	\begin{split}	
	&\int_L dJ \dfrac{\tilde{f}_0'(J)}{\Omega(J)} \int_{-\pi}^{\pi} \cos^2 q(\theta, J) d\theta \\
        & = \int_{L} dJ \int_{-\pi}^{\pi} 
        \dfrac{d \hat{f}_0}{d\mathcal{E}} (\mathcal{E}(J)) 
        \cos^{2}q(\theta, J) ~d\theta\\
	& = \iint_\mu  \dfrac{d \hat{f}_0}{d\mathcal{E}} (\mathcal{E}(J)) 
        \cos^2 q(\theta, J) ~d\theta dJ\\
	& = \iint_{\mu} \dfrac{1}{p} \dfrac{\partial f_0}{\partial p} (q,p)\cos^2 q ~ dq dp. 
	\end{split}
\end{equation}
In the course of analysis, we have used the fact 
that the transformation $(\theta, J) \mapsto (q,p)$ is canonical. 
We note that 
$d \hat{f}_0/d\mathcal{E}$ is assumed to be continuous in $J$, 
and hence the integration with respect to $J$ is taken along on the real $J$-axis. 
Then, by using \eqref{eq:intg} and \eqref{eq:intf}, the second equality in 
\eqref{eq:int1Dx0} is derived. 
Equations \eqref{eq:aaD0}, \eqref{eq:intDx0}, and \eqref{eq:int1Dx0} 
are put together to show the relation $D_x(0) = \mathcal{I}[f_0]$.

So far we have investigated the stability against perturbations 
in the direction parallel to the order parameter $\vec{M}_0 = (M_0, 0)^T$. 
We proceed to look into the stability against a perturbation 
in the direction 
perpendicular to the order parameter $\vec{M}_0$. 
The dispersion relation corresponding to 
the direction perpendicular to the order parameter $\vec{M}_0$ is expressed as 
\begin{equation}
	\label{eq:inhom-disp-perp}
	\begin{split}
		D_y(\omega) &= 1 + 2\pi \sum_{m \in \mathbb{Z}} \int_L 
		\frac{m {\tilde{f}_0}'(J)}{m\Omega(J) - \omega}|S^m(J)|^2 dJ
		=0, 
	\end{split}
\end{equation} 
for ${\rm Im} \omega > 0$, where 
\begin{equation}
	S^m(J) \equiv \frac{1}{2\pi}\int_{-\pi}^{\pi} \sin q(\theta, J) e^{-im\theta} d\theta.
\end{equation}
We note that $S^0(J) = 0$.  
In fact, $\sin q (\theta, J)$ is expressed as 
\begin{equation}
	\label{eq:sink}
	\begin{split}
	&\sin q\left(\theta, J(k)\right) 
	= 
	\begin{cases}
		\vspace{5mm}
		\displaystyle
		2k \sn\left(\frac{2K(k)}{\pi} \theta, k\right)
		\dn\left(\frac{2K(k)}{\pi} \theta, k\right), & k< 1, \\
		\displaystyle
		2 \sn \left(\frac{K(1/k)}{\pi} \theta, \frac{1}{k} \right)
		\cn \left(\frac{K(1/k)}{\pi} \theta, \frac{1}{k} \right),& k>1,
	\end{cases}
	\end{split}
\end{equation}
it is odd with respect to $\theta$ for 
all $J$ \cite{MAIS72}, 
so that one has $S^0(J) = 0$.

Following the same procedure as that for proving the relation $D_x(0) = \mathcal{I}[f_0]$ 
and taking into account the relation $S^0(J) = 0$, we obtain 
\begin{equation}
	D_y(0) = 
	1 + \iint_\mu 
	\frac{1}{p} \dfrac{\partial f_0}{\partial p}(q,p) \sin^2q ~ dqdp .
\end{equation} 
If $D_y(0) \geq 0$, 
there is no eigenmode which brings about the instability 
in a direction perpendicular to the order parameter $\vec{M}_0$. 
Actually, the equality, $D_y(0)= 0$, 
is satisfied for any stationary solution subject to 
the assumptions 
in Prop. \ref{prop2} with \eqref{e:1bodyE}, 
which is proved as follows \cite{AC10}; 
\begin{equation}
	\label{eq:Dy00}
	\begin{split}
	&\iint_\mu 
	\frac{1}{p} \dfrac{\partial f_0}{\partial p}(q,p) \sin^2q ~ dqdp\\
	&= 
	\iint_\mu 
	\frac{d \hat{f}_0}{d\mathcal{E}} \left(\mathcal{E}(q,p) \right) \sin^2 q ~ dqdp\\ 
	&= 
	\frac{1}{M_0}\iint_\mu \frac{\partial f_0}{\partial q}(q,p) \sin q ~dqdp \\ 
	&= 
	-\frac{1}{M_0} \iint  f_0(q, p) \cos q~dq dp= -1. 
\end{split}
\end{equation}

Since $D_x(0) = \mathcal{I}[f_0]$ and $D_y(0) = 0$, 
we have obtained 
the spectral stability criterion \eqref{eq:inhom-crit} 
for the spatially inhomogeneous solutions to the Vlasov equation. 

It is to be remarked that any spectrally stable solution 
which are spatially inhomogeneous are neutrally spectrally stable, 
since there is an 
embedded eigenvalue 0 which comes from $D_y(0)=0$. 

To compute $D_x(0)$ 
or the right-hand side of \eqref{eq:inhom-crit}, 
we should express $C^0(J)$ 
in terms of known functions. 
On using the explicit expression of $\Omega(J)$ and $C^0(J)$ given respectively in 
Appendix B of \cite{BOY10} and Appendix C of \cite{SO12}, 
$D_x(0)$ is described explicitly as 
\begin{equation}
	\label{eq:Dx0}
	\begin{split}
	&D_x(0) =  1 + \iint_{\mu} \frac{1}{p} \frac{\partial f_0}{\partial p} (q, p) 
	\cos^2q~ dqdp \\
	&- \frac{4}{\sqrt{M_0}} \int_0^1 K(k) \left[\frac{2 E(k)}{K(k)}-1\right]^2 \bar{f}_0'(k)~dk \\
	&- \frac{4}{\sqrt{M_0}} \int_1^\infty \frac{K(1/k)}{k} 
	\left[\frac{2k^2 E(1/k)}{K(1/k)}+1-2k^2\right]^2 \bar{f}_0'(k)~dk,
	\end{split}
\end{equation}
where  
$\bar{f}_0\left(k\right) \equiv \tilde{f}_0\left(J(k)\right)$, 
and where $E(k)$ is the complete elliptic integral 
of the second kind \cite{MAIS72}.

\subsection{Stationary states realized as 
critical points of some invariant functionals} 
\label{subsec:opt}
We will give a necessary and sufficient condition 
of the formal stability of a stationary state. 
To look into the formal stability, 
we introduce invariant functionals. 

The Vlasov dynamics satisfies 
the normalization condition, 
\begin{equation}
\label{eq:normalization}
	\mathcal{N}[f] 
	\equiv \iint_\mu f (q,p)~dq dp = 1, 
\end{equation}
the momentum conservation law, 
\begin{equation}
	\label{eq:momentum}
	\mathcal{P}[f] \equiv \iint_\mu p f(q,p)~dqdp = 0, 
\end{equation}
and the energy conservation law, 
\begin{equation}
	\label{eq:energy}
	\mathcal{U}[f] 
	\equiv \iint_\mu \frac{p^2}{2} f(q,p)~dq dp - 
	\frac{1}{2} \left(M_x[f]^2 + M_y[f]^2 \right) = U,  
\end{equation}
where $U$ is a fixed value. 
The Vlasov dynamics additionally has an infinite number of 
Casimir invariants denoted by 
\begin{equation}
	\label{eq:cas-func}
	\mathcal{S}[f] = \iint_\mu s\left(f(q,p) \right)~dq dp. 
\end{equation}
We here assume that $s$ is a strictly concave and 
twice differentiable function 
defined for the non-negative real numbers. 

We will look into the formal stability of the 
stationary solution 
realized as the critical point of \eqref{eq:cas-func} 
under constraints 
\eqref{eq:normalization}, 
\eqref{eq:momentum}, and \eqref{eq:energy}.  
A critical point $\tilde{f}_0(J)$ is a 
solution to the variational equation 
\begin{equation}
	\label{eq:vareq}
	\delta \mathcal{F} = \delta (\mathcal{S}  - \beta \mathcal{U}- \alpha \mathcal{N}) = 0, 
\end{equation}
which is written out as 
\begin{equation}
	\label{eq:solvar}
	s'\left(\tilde{f}_0(J)\right) = \beta \mathcal{E}(J) + \alpha, 
\end{equation}
where $\alpha$ and $\beta$ are Lagrangian multipliers. 
Since $s(x)$ is a strictly concave differentiable function defined on $x \geq 0$, 
its derivative $s'(x)$ is strictly decreasing on $x \geq 0$, and the inverse function 
$\left(s'\right)^{-1}(y)$ exists and is strictly decreasing on the range of the function $s'$. 
We are then allowed to put the solution $\tilde{f}_0(J)$ 
to the variational equation \eqref{eq:vareq} in the form
\begin{equation}
	\label{eq:solsolvar}
	\hat{f}_0(\mathcal{E}) = \tilde{f}_0\left(J\left(\mathcal{E}\right)\right) 
	= \left(s'\right)^{-1}\left(\beta \mathcal{E}+ \alpha\right). 
\end{equation}

The parameter $\beta$ is positive \cite{AC10}. 
To see this, we assume that $\beta$ were not positive.
(i) When $\beta < 0$, from \eqref{eq:solsolvar}, 
the function $\hat{f}_0\left(\mathcal{E}\right)$ 
is strictly increasing with respect to $\mathcal{E}$, 
so that the function $\tilde{f}_0\left(J\right)$ 
is strictly increasing with respect to $J$.
(ii) When $\beta = 0$, $\hat{f}_0(\mathcal{E})$ is a constant for the whole $\mathcal{E}$, 
so that $\tilde{f}_0(J)$ is a constant for the whole $J$. 
In these cases, 
the integral $\int_L \tilde{f}_0(J)~dJ$ diverges, 
and hence $\tilde{f}_0(J)$ 
cannot be a probabilistic density function. 
Hence, parameter $\beta$ must be positive. 
In the case $\beta > 0$,  
$\tilde{f}_0(J)$ can be a probabilistic density function.

Since $\beta$ is shown to be positive, and 
since $s$ is strictly concave, 
a solution \eqref{eq:solsolvar} 
to the variational equation \eqref{eq:vareq} 
is a stationary solution to the Vlasov equation satisfying 
$d \hat{f}_0/ d\mathcal{E} < 0$ 
and $d \tilde{f}_0/ d J < 0$. 

\subsection{Formal stability criterion in the most refined form}
\label{subsec:inhom-formal}

In this section, we look into 
the most refined formal stability of 
the spatially inhomogeneous stationary solution $f_0$ 
which is a critical point of the functional \eqref{eq:cas-func} 
under the constraint conditions \eqref{eq:normalization}, 
\eqref{eq:momentum}, and \eqref{eq:energy}.   
To start with, we note that $C^n(J) = C^{-n}(J)$. 
In fact, from $\sn(u, k) = -\sn(-u, k)$ \cite{MAIS72} and \eqref{eq:cosk}, one has that 
$\cos q(\theta, J)$ is even with respect to $\theta$, so that 
$C^n(J)$ is shown to be real from the definition \eqref{eq:cn} and 
$C^n(J) = C^{-n}(J)$, and further $|C^n(J)|^2 = C^n(J)^2$ .

We derive the formal stability criterion for spatially 
inhomogeneous solutions on the basis of the following claim. 
\begin{claim} 
\label{claim1}
A solution $\tilde{f}_0(J)$ to the variational equation \eqref{eq:vareq} is formally stable, 
if and only if the second-order variation 
of the functional $\mathcal{F} = \mathcal{S}-\beta \mathcal{U}-\alpha \mathcal{N}$ 
is negative definite at $\tilde{f}_0$ under the constraint of the Casimir invariants. 
That is, $\delta^2 \mathcal{F}\left[f_0\right][\delta f, \delta f] < 0$ 
for any non-zero variation $\delta f$ leaving invariant the functional of the form \eqref{eq:en-cas0} 
up to first order for any function $Q$. 
\end{claim}

To investigate the condition $\delta^2 \mathcal{F}\left[f_0\right][\delta f, \delta f] <0$, 
we start by putting the function $\gamma$ as
\begin{equation}
	\label{eq:gammaJ}
	\gamma(J) = \frac{\beta}{s'' \left(\tilde{f}_0(J)\right) } 
	= \frac{\tilde{f}_0'(J)}{\Omega(J)}
	= \dfrac{d \hat{f}_0}{d\mathcal{E}}\left(\mathcal{E}(J)\right).
\end{equation}
Then 
the second-order variation of $\mathcal{F}$ is described as 
\begin{equation}
	\label{eq:second-var}
	\begin{split}
		\delta^2\mathcal{F}\left[\tilde{f}_0\right]\left[\delta \tilde{f}, \delta \tilde{f}\right]
		&=
		\iint_\mu \frac{\beta}{\gamma(J)} \delta \tilde{f}(\theta, J)^2 ~d\theta dJ \\
		&+ \beta \left[
		\iint_\mu \cos q(\theta,J) ~\delta \tilde{f}(\theta, J)~d\theta dJ 
		\right]^2\\
		&+ \beta \left[
		\iint_\mu \sin q(\theta,J) ~\delta \tilde{f}(\theta, J)~d\theta dJ 
		\right]^2.
	\end{split}
\end{equation}

On account of the constraints of the Casimir invariants \eqref{eq:en-cas0} up to first order, 
the perturbation should satisfy the constraint 
\begin{equation}
	\label{eq:en-cas-pert}
	\begin{split}
		\mathcal{Q}[f_0 + \delta f] - \mathcal{Q}[f_0]  
		&=
		\iint_\mu \mathcal{Q}'\left(f_0(q,p) \right) \delta f (q, p) ~dqdp 
		\\
		&= 
		\int_L dJ~Q'\left(\tilde{f}_0(J)\right)
		\int_{-\pi}^{\pi} 
		\delta \tilde{f}(\theta, J) ~d\theta \\
		&= 0.
	\end{split}
\end{equation} 
Since $Q$ is chosen arbitrarily, 
we can look on $Q' \left(\tilde{f}_0(J)\right)$ as 
a function of $J$ (or $\mathcal{E}(J)$) chosen arbitrarily. 
We are then allowed to restrict perturbations to those satisfying 
\begin{equation}
	\label{eq:en-cas-equi}
	 \int_{-\pi}^{\pi} \delta \tilde{f}(\theta, J) ~d \theta = 0, 
	 \quad \forall J.
\end{equation}
We now divide the perturbation $\delta \tilde{f} (\theta, J)$ into 
even and odd parts with respect to $\theta$, 
\begin{equation}
	\label{eq:odd-even}
	\delta \tilde{f}(\theta, J) = 
	\delta_{\rm e} \tilde{f}(\theta, J) + 
	\delta_{\rm o} \tilde{f}(\theta, J),
\end{equation}
where
\begin{equation}
	\begin{split}
		\delta_{\rm e}\tilde{f}(\theta, J)&=
		\frac{1}{2} \left( 
		\delta \tilde{f}(\theta, J)+\delta \tilde{f}(-\theta, J)
		\right), 
		\\
		\delta_{\rm o}\tilde{f}(\theta, J)&=
		\frac{1}{2} \left( 
		\delta \tilde{f}(\theta, J)-\delta \tilde{f}(-\theta, J)
		\right).
	\end{split}
\end{equation}
When $\delta \tilde{f}$ in the functional \eqref{eq:second-var} 
is replaced by \eqref{eq:odd-even}, 
the functional \eqref{eq:second-var} is arranged as 
\begin{equation}
	\label{eq:second-var2}
	\begin{split}
		\delta^2\mathcal{F}\left[\tilde{f}_0\right]
		&\left[\delta \tilde{f}, \delta \tilde{f}\right]
		=
		\iint_\mu \frac{\beta}{\gamma(J)} \delta_{\rm e} \tilde{f}(\theta, J)^2 ~d\theta dJ \\
		&+ \beta \left[
		\iint_\mu \cos q(\theta,J) ~\delta_{\rm e} \tilde{f}(\theta, J)~d\theta dJ 
		\right]^2\\
		&+ \iint_\mu \frac{\beta}{\gamma(J)} \delta_{\rm o} \tilde{f}(\theta, J)^2 ~d\theta dJ\\
		&+ \beta \left[
		\iint_\mu \sin q(\theta,J) ~\delta_{\rm o} \tilde{f}(\theta, J)~d\theta dJ 
		\right]^2\\
		&=\delta^2\mathcal{F}\left[\tilde{f}_0\right]
		\left[\delta_{\rm e} \tilde{f}, \delta_{\rm e} \tilde{f}\right]+
		\delta^2\mathcal{F}\left[\tilde{f}_0\right]
		\left[\delta_{\rm o} \tilde{f}, \delta_{\rm o} \tilde{f}\right], 
	\end{split}
\end{equation}
where we have used the fact that
\begin{equation}
	\label{eq:mxomye}
	M_x\left[\delta_{\rm o} \tilde{f}\right] = 0, \quad M_y\left[\delta_{\rm e} \tilde{f}\right] = 0,  
\end{equation}
which come from the fact that 
$\cos q(\theta, J)$ (resp. $\sin q(\theta, J)$)
is even (resp. odd) with respect to $\theta$ 
on account of \eqref{eq:cosk} (resp. $\eqref{eq:sink}$). 
Equation \eqref{eq:second-var2} means that 
$\delta_{\rm e} \tilde{f}$ and $\delta_{\rm o}\tilde{f}$ are not 
coupled in \eqref{eq:second-var2}. 
As for the second term in the right-hand side of the last equality 
in \eqref{eq:second-var2}, we recall that 
spatially inhomogeneous stationary solutions are 
already known to be neutrally formally stable against 
a perturbation $\delta_{\rm o} \tilde{f}$ whose direction is 
perpendicular to the direction of the order parameter $\vec{M}_0$, 
as is shown in \cite{AC10}. 
This fact is consistent with the fact that the order parameter may rotate
if an arbitrarily small external field is turned on 
perpendicularly to the order parameter \cite{SO12}. 
We do not take into account  this rotation 
as long as we treat a formal stability of the stationary solution $f_0$,
as we mentioned in Sec. \ref{sec:LVE}.   
On account of \eqref{eq:mxomye}, 
we are now left with the analysis of, 
$\delta^2\mathcal{F}\left[\tilde{f}_0\right]
\left[\delta_{\rm e} \tilde{f}, \delta_{\rm e} \tilde{f}\right]$, 
the integrals in \eqref{eq:second-var2} 
for the even part $\delta_{\rm e} \tilde{f}$ whose direction is 
parallel to the order parameter $\vec{M}_0$.

In what follows, we prove the proposition: 
\begin{proposition}
\label{prop-f-stab}
Let $f_0$ be a solution to the variational equation \eqref{eq:vareq}. 
The inequality 
\begin{equation}
	\label{eq:f-stab}
	\mathcal{I}[f_0] = D_x(0) > 0
\end{equation}
is equivalent to the condition
\begin{equation}
	\delta^2 \mathcal{F}\left[\tilde{f}_0\right]
	\left[\delta_{\rm e} \tilde{f}, \delta_{\rm e} \tilde{f}\right] < 0
\end{equation} 
for any $\delta_{\rm e} \tilde{f} \neq 0$ 
under the constraint \eqref{eq:en-cas-equi}. 
Therefore, the inequality \eqref{eq:f-stab} is a 
necessary and sufficient condition for the  
formal stability of $\tilde{f}_0$.
\end{proposition}

In the situation stated so far, 
the second order variation \eqref{eq:second-var} is put in the form
\begin{equation}
	\label{eq:second-var-e}
	\begin{split}
		\delta^2\mathcal{F}\left[\tilde{f}_0\right]
		\left[\delta_{\rm e} \tilde{f}, \delta_{\rm e} \tilde{f}\right]&=
		\iint_\mu \frac{\beta}{\gamma(J)} \delta_{\rm e} \tilde{f}(\theta, J)^2 ~d\theta dJ \\
		+ &\beta \left[
		\iint_\mu \cos q(\theta,J) \delta_{\rm e} \tilde{f}(\theta, J)~d\theta dJ 
		\right]^2. 
	\end{split}
\end{equation}

We first show that a non-zero $\delta_{\rm e} \tilde{f}$ satisfying 
$M_x\left[\delta_{\rm e} \tilde{f}\right] = 0$ does not bring about 
the formal instability. 
Indeed, \eqref{eq:second-var-e} becomes 
\begin{equation}
		\delta^2\mathcal{F}\left[\tilde{f}_0\right]
		\left[\delta_{\rm e} \tilde{f}, \delta_{\rm e} \tilde{f}\right]=
		\iint_\mu \frac{\beta}{\gamma(J)} \delta_{\rm e} \tilde{f}(\theta, J)^2 ~d\theta dJ, 
\end{equation}
and is negative 
since $\gamma(J) < 0$ and 
$\beta > 0$, as was mentioned in Sec. \ref{subsec:opt}.

We proceed to perform the stability analysis 
with the constraint condition  
\begin{equation}
\label{eq:const-m}
	M_x\left[\delta_{\rm e} \tilde{f}\right] =\iint_\mu \cos q(\theta, J) \delta_{\rm e} \tilde{f}(\theta, J)~d\theta dJ
	=1 .
\end{equation}
We note that the value of $M_x\left[\delta_{\rm e}\tilde{f}\right]$ can be chosen arbitrary, 
because this value changes only the scaling of \eqref{eq:second-var-e} and 
does not change the sign of \eqref{eq:second-var-e}. 
We expand the perturbation $\delta_{\rm e} \tilde{f}$ 
into the Fourier series in $\theta$,
\begin{equation}
	\label{eq:even-f}
	\begin{split}
		\delta_{\rm e}\tilde{f}(\theta, J) = 
		\sum_{n\neq 0} \hat{f}^{\rm e}_n(J) e^{i n \theta}, \quad
		\hat{f}^{\rm e}_n (J) &= \hat{f}^{\rm e}_{-n}(J).   
	\end{split}
\end{equation}
We note that the 0-th Fourier mode vanishes 
thanks to the constraint condition \eqref{eq:en-cas-equi}. 
Substituting \eqref{eq:even-f} into \eqref{eq:second-var-e}, 
we obtain the functional in $\left\{\hat{f}^{\rm e}_n\right\}_{n \neq 0}$
\begin{widetext}
\begin{equation}
	\label{eq:functional-G}		
	\begin{split}
		\mathcal{G}_{\rm e}\left[\left\{\hat{f}^{\rm e}_n\right\}_{n \neq 0}\right] 
		&\equiv 
		\frac{1}{2\pi} \delta^2\mathcal{F}\left[\tilde{f}_0\right]
		\left[\delta_{\rm e} \tilde{f},\delta_{\rm e} \tilde{f}\right] \\
		&=
		\sum_{n \neq 0} 
		\int_L  
		 \frac{\beta}{\gamma(J)} \hat{f}^{\rm e}_n(J)^2 ~dJ
		 + 2 \pi \beta 
		 \left(\sum_{m\neq 0} \int_L 
		 C^m(J') \hat{f}^{\rm e}_m (J')~dJ'
		 \right)^2 .
	\end{split}
\end{equation}
We look for a critical point of $\mathcal{G}_{\rm e}$ under the constraint \eqref{eq:const-m} 
which is rewritten in terms of $\left\{\hat{f}^{\rm e}_n\right\}_{n \neq 0}$ as 
\begin{equation}
	\label{eq:const-c}
	\mathcal{M}_x\left[\left\{\hat{f}^{\rm e}_m\right\}_{m \neq 0}\right]  
	\equiv 2\pi\sum_{m \neq 0}\int_L C^m(J)  \hat{f}^{\rm e}_m (J)~dJ = 1.
\end{equation}

The functional $\mathcal{G}_{\rm e}\left[\left\{\hat{f}^{\rm e}_m\right\}_{m \neq 0}\right]$ 
takes a critical value under the constraint condition \eqref{eq:const-c} if
\begin{equation}
	\label{eq:critical-eq}
	\delta_n \mathcal{G}_{\rm e}\left[\left\{\hat{f}^{\rm e}_m\right\}_{m \neq 0}\right] - 
	\eta\delta_n \mathcal{M}_x\left[\left\{\hat{f}^{\rm e}_m\right\}_{m \neq 0}\right] = 0, \quad 
	n \in \mathbb{Z} \setminus \{0\},
\end{equation}
where $\eta$ is a Lagrangian multiplier, and $\delta_n \mathcal{G}_{\rm e}$ is defined by
\begin{equation}
	\begin{split}
		\delta_n \mathcal{G}_{\rm e}\left[\left\{\hat{f}^{\rm e}_m\right\}_{m \neq 0}\right] 
		&\equiv 
		 \mathcal{G}_{\rm e}\left[\left\{\hat{f}^{\rm e}_m + \delta \hat{f}^{\rm e}_n \delta_{mn}\right\}_{m \neq 0}\right] 
		 - \mathcal{G}_{\rm e}\left[\left\{\hat{f}^{\rm e}_m\right\}_{m \neq 0}\right] \\
		&= 
		2\beta \int_L \delta \hat{f}^{\rm e}_n(J) 
		\left[ \frac{\hat{f}^{\rm e}_n (J)}{\gamma(J)}  
		+ 2 \pi C^n(J) \left( \sum_{m \neq 0} \int_L C^m(J') \hat{f}^{\rm e}_m(J')~dJ'\right)
		\right]~dJ, 
	\end{split}
\end{equation}
where $\delta_{mn}$ is the Kronecker delta. 
Hence, Eq. \eqref{eq:critical-eq} results in 
\begin{equation}
	\label{eq:hatf}
	\begin{split}
	\hat{f}^{\rm e}_n(J)
	&= - \left(2 \pi  \sum_{m \neq 0} \int_L C^m(J') \hat{f}_m^{\rm e}(J')~dJ'\right)C^n(J) \gamma(J)
		+\frac{\pi\eta}{\beta} C^n(J) \gamma(J) 
	= - \xi C^n(J)\gamma(J), 	
	\end{split}
\end{equation}
for all $n \in \mathbb{Z}\setminus\{0\}$, 
where 
we have used \eqref{eq:const-c} and put $\xi \equiv 1 - \pi \eta/\beta$. 
Substituting \eqref{eq:hatf} into \eqref{eq:const-c}, 
we obtain the value of $\xi$ as  
\begin{equation}
	\xi = \frac{-1}{\displaystyle 2\pi \sum_{m \neq 0} \int_L C^m(J)^2  \gamma(J)~dJ}.
\end{equation}
A non-vanishing critical point 
$\left\{\hat{f}^{\rm e, m}_n\right\}_{n \in \mathbb{Z} \setminus \{0\}}$ is therefore given by
\begin{equation}
	\label{eq:nvc}
	\hat{f}^{\rm e, m}_n(J) = 
	\frac{C^n(J) \gamma(J)}{\displaystyle 2\pi \sum_{m \neq 0} \int_L C^m(J')^2 \gamma(J')~dJ'}, 
\quad n \in \mathbb{Z} \setminus \{0\}.  
\end{equation}
Substituting \eqref{eq:nvc} into 
\eqref{eq:functional-G}, 
we obtain 
\begin{equation}
	\label{eq:gmax}
	\begin{split}
	\mathcal{G}_{\rm e}\left[\left\{\hat{f}^{\rm e, m}_n\right\}_{n \neq 0}\right] 
		=&
		\frac{\beta}{\displaystyle 4\pi^2\sum_{m \neq 0} \int_L C^m(J)^2\gamma(J) ~dJ} + \frac{\beta}{2\pi} 
		\\
		=& 
		\frac{\beta}{\displaystyle 4\pi^2 \sum_{m \neq 0} \int_L C^m(J)^2 \gamma(J) ~dJ}
		\times\left[ 
			1 + 2\pi \sum_{l \neq 0} \int_L \frac{\tilde{f}_0'(J)}{\Omega(J)} C^l(J)^2~dJ
		\right], 
	\end{split}
\end{equation}
where we have used \eqref{eq:gammaJ}.
\end{widetext}

Since $\gamma(J) < 0$, and since $C^n(J) \neq 0$ for 
some $J$ and $n\in \mathbb{Z} \setminus \{0\}$, 
we have
\begin{equation}
	\label{eq:ineq-up}
	\begin{split}
		\sum_{m\neq 0} \int_L C^m(J)^2 \gamma(J)~ dJ < 0 .
	\end{split}
\end{equation} 
It then follows, from \eqref{eq:gmax} 
along with \eqref{eq:ineq-up} and 
the positivity of $\beta$ which has been shown at the end of Sec. \ref{subsec:opt}, 
that the quadratic form 
\eqref{eq:functional-G} is negative definite if and only if the inequality 
\begin{equation} 
	D_x(0) = 1 + 2\pi \sum_{l \neq 0} \int_L \frac{\tilde{f}_0'(J)}{\Omega(J)} C^l(J)^2~dJ > 0
\end{equation}
is satisfied.
We hence conclude that 
the inequality \eqref{eq:f-stab} is a 
necessary and sufficient condition for formal stability. 
Once the criterion \eqref{eq:f-stab} is obtained, 
we no longer have to seek an infinite number of Lagrangian multipliers 
to get the most refined formal stability criterion given in \cite{AC10}.
The formal stability criterion \eqref{eq:f-stab} 
is stronger than 
the condition that $\tilde{f}_0(J)$ is spectrally stable 
in the sense that the equality in 
\eqref{eq:inhom-crit} is not allowed. 

\begin{remark}
We have shown that the stability of $f_0$ is 
determined by the sign of $\mathcal{I}[f_0] = D_x(0)$. 
Further the value of the positive $\mathcal{I}[f_0]$ 
is thought to express a strength of stability of $f_0$ since 
the zero-field isolated-susceptibility $\chi$ is derived as 
\begin{equation}
	\label{eq:chi0}
	\chi = \frac{1-D_x(0)}{D_x(0)} = \frac{1-\mathcal{I}\left[f_0\right]}{\mathcal{I}\left[f_0 \right]},  
\end{equation}
with the linear response theory based on 
the Vlasov equation \cite{SO12}. 
Equation \eqref{eq:chi0} implies that 
stability of a stationary state $f_0$ becomes 
stronger as $\mathcal{I}[f_0]$ becomes larger, 
since $\mathcal{I}[f_0] = D_x(0) \leq 1$. 
The last inequality is derived as follows. 
As we mentioned in Sec. \ref{subsec:inhom-spec}, 
the function $D_x (i \omega_{\rm i})$ is strictly increasing 
and continuous with respect to $\omega_{\rm i} \geq 0$, 
and $D_x (i\omega_{\rm i}) \to 1$, as $\omega_{\rm i} \to \infty$. 
\end{remark}

\subsection{Observation of the criteria} 
\label{subsec:water-bag} 
\begin{figure}[tb]
	\begin{center}
	        \includegraphics[width=8.5cm]{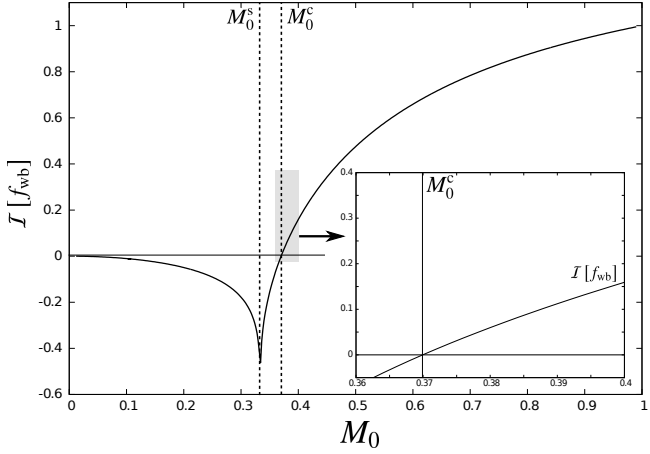}\\
	        \caption{
		Plot of $\mathcal{I}\left[f_{\rm wb}\right]$ as 
		a function of $M_0$: 
		$\mathcal{I}\left[f_{\rm wb} \right]>0$ for $M_0 > M_0^{\rm c}$. 
		The critical value 
		$M_0^{\rm c} \simeq 0.369942$. 
		The edge of the water-bag $\mathcal{E}(q,p) = \mathcal{E}^\ast$ 
		coincides with the separatrix when 
		$M_0 = M_0^{\rm s} \simeq 0.33$. 
		}
	        \label{fig-example}
	    \end{center}
\end{figure}

\begin{figure}[bt]
	\begin{center}
		\includegraphics[width=8.0cm]{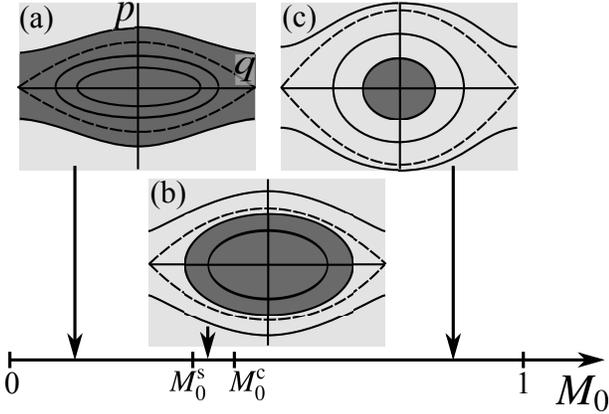}
		\caption{
		Gray rectangles are $\mu$-spaces for each $M_0$. 
		The curves in $\mu$-spaces are iso-$\mathcal{E}$ lines, 
		and the broken curves are separatrices. 
		On the dark gray region in each $\mu$-space, 
		the water-bag distribution takes non-zero value, $\eta_0$.  
		}
		\label{fig-illust}
	 \end{center}
\end{figure}

Let us observe what kinds of stationary states 
are likely to be stable 
through the stability analysis for a family of the stationary 
water-bag distributions \cite{BOY10}, 
\begin{equation}
	\label{eq:wb}
	\begin{split}
		f_{\rm wb}(q,p) &= \eta_0 \Theta\left(\mathcal{E}^\ast - \mathcal{E}(q,p) \right), \\
		\mathcal{E}(q,p) &= \frac{p^2}{2} - M_0 \cos q, 
	\end{split}
\end{equation}
where $\Theta$ is the Heaviside step function. 
Though the water-bag distributions \eqref{eq:wb} do not satisfy assumptions in Prop. \ref{prop2}, 
they make it possible to observe the stability visually. 
Let us put $k^\ast = \sqrt{(\mathcal{E}^\ast + M_0)/(2M_0)}$. 
For each fixed $M_0$, the two parameters 
$\eta_0$ and $\mathcal{E}^\ast$ are determined by 
the normalization condition 
\begin{equation}
	\label{eq:n-wb}
	\begin{split}
		1 
		&= \iint_\mu f_{\rm wb}(q,p) ~dqdp \\
		&=
		\begin{cases}
			16 \eta_0 \sqrt{M_0} \left(E(k^\ast)-(1-{k^\ast}^2) K(k^\ast) \right), \quad& k^\ast < 1,\\
			16 \eta_0 \sqrt{M_0} k^\ast E(1/k^\ast), & k^\ast > 1,
		\end{cases} 
	\end{split}
\end{equation}
and the self-consistent equation, 
\begin{widetext}
\begin{equation}
	\label{eq:sc-wb}
	\begin{split}
		M_0 &= \iint_{\mu} \cos q f_{\rm wb}(q,p) ~dqdp\\
		&= 
		\begin{cases} \displaystyle
			1- \frac{2}{3}
			\frac{(2-{k^\ast}^2)E(k^\ast) - (2-2{k^\ast}^2) K(k^\ast)}{E(k^\ast) - (1-{k^\ast}^2)K(k^\ast)}, \, &k^\ast < 1,\\
		\displaystyle
			\frac{2{k^\ast}^2-1}{3} - \frac{2{k^\ast}^2-2}{3} \frac{K(1/k^\ast)}{E(1/k^\ast)}, \, &k^\ast > 1.
		\end{cases}
	\end{split}
\end{equation}
For the water-bag distribution \eqref{eq:wb}, 
we are able to compute $\mathcal{I}\left[f_{\rm wb}\right]$ explicitly by using 
equations 
\begin{equation}
	\label{eq:dp-de}
	\begin{split}
		\frac{1}{p} \frac{\partial f_{\rm wb}}{\partial p}(q,p) 
		&= 
		\frac{d \hat{f}_{\rm wb}}{d \mathcal{E}} \left(\mathcal{E}(q,p) \right) 
		= -\eta_0 \delta\left(\mathcal{E}^\ast- \mathcal{E}(q,p)\right), 
	\end{split}
\end{equation}
and  
\begin{equation}
	\label{eq:dk-de}
	\begin{split}
		\frac{d \bar{f}_0}{d k}(k) 
		&= 4 M_0 k \frac{d \hat{f}_0}{d \mathcal{E}} \left(\mathcal{E}(k) \right)
		= -4 M_0 k \eta_0 \delta\left(\mathcal{E}^\ast- \mathcal{E}(k)\right),
	\end{split}
\end{equation}
and by using Eq. \eqref{eq:Dx0}. 
Then $\mathcal{I}\left[ f_{\rm wb}\right]$ is written as 
\begin{equation}
	\label{eq:Iwb}
	\begin{split}
		\mathcal{I}\left[f_{\rm wb}\right] =&
		\begin{cases} 
		\displaystyle
			 1 +  
			 \frac{\left(8{k^\ast}^2 -4\right) E(k^\ast) + \left(1-4{k^\ast}^2\right)K(k^\ast)}
			 {12M_0\left[E(k^\ast) -\left(1-{k^\ast}^2\right)K(k^\ast)\right]} 
			+  \frac{\left(2E(k^\ast)-K(k^\ast)\right)^2}
			{4M_0K(k^\ast) \left[ E(k^\ast) - (1-{k^\ast}^2) K(k^\ast)\right]}, ~ &k^\ast < 1, \\
		\displaystyle
			1+ \frac{(8{k^\ast}^4-4{k^\ast}^2) E(1/k^\ast)-(8 {k^\ast}^4 - 8{k^\ast}^2 + 3)K(1/k^\ast)}
			{12 M_0 {k^\ast}^2 E(1/k^\ast)}
			+ \frac{\left(2{k^\ast}^2 E(1/k^\ast) + (1-2{k^\ast}^2) K(1/k^\ast)\right)^2}
			{4 M_0 k^\ast K(1/k^\ast) E(1/k^\ast)} ,~& k^\ast > 1.
		\end{cases}
	\end{split}
\end{equation}
\end{widetext}
Since $\mathcal{E}^\ast$ and $k^\ast$ is determined by $M_0$, 
then $\mathcal{I} \left[f_{\rm wb} \right]$ in \eqref{eq:Iwb} 
can be looked on as a function of $M_0$ and 
it is plotted in Fig. \ref{fig-example}. 
According to this graph, 
the water-bag $f_{\rm wb}$ is formally 
(resp. spectrally) 
stable 
when $M_0 > M_0^{\rm c} \simeq 0.369942$ 
(resp. when $M_0 \geq M_0^{\rm c}$). 
The critical value $M_0^{\rm c}\simeq 0.369942$ is obtained by solving 
the self-consistent equation \eqref{eq:sc-wb} and 
$\mathcal{I}\left[f_{\rm wb}\right] = 0$ simultaneously, and 
it is close to the estimation $M_0^{\rm c} \simeq 0.37$ in \cite{BOY10}. 
The water-bag distribution with large $M_0$ tends to be stable, 
and the stability of it tends to be strong, 
since $\mathcal{I}\left[f_{\rm wb}\right]$ is monotonically increasing with 
respect to $M_0$, when 
$M_0 > M_0^{\rm s} \simeq 0.33$. 
We illustrate it in Fig. \ref{fig-illust}. 
The water-bag distributions illustrated in panels (a) or (b) are unstable, 
and one illustrated in panel (c) is stable. 

\section{Comparison with the Canonical formal stability}
\label{sec:example}

Let us compare the formal stability criterion \eqref{eq:f-stab}, 
$\mathcal{I}[f_0] = D_x(0) >0$, with  
\textit{the canonical formal stability} criterion given in \cite{AC10}. 
We start with a brief review of the canonical formal stability. 

\subsection{The canonical formal stability}
\begin{claim}[\cite{AC10}]
A solution $f_0$ to the variational equation \eqref{eq:vareq} is called 
canonically formally stable 
against any perturbation $\delta_{\rm e} f$ whose direction is parallel to 
the order parameter $\vec{M}_0 = (M_0, 0)^T$, if and only if the second order variation 
of the functional $\mathcal{F} = \mathcal{S}- \beta \mathcal{E} -\alpha \mathcal{N}$ 
at $f_0$, $\delta^2 \mathcal{F}[f_0]$, subject to the normalization condition 
is negative definite, i.e.,
\begin{equation}
	\label{eq:second-var-c}
	\delta^2 \mathcal{F}[f_0] \left[\delta_{\rm e} f, \delta_{\rm e} f\right] < 0, 
\end{equation}
for all $\delta_{\rm e} f \neq 0$ satisfying 
\begin{equation}
	\iint_\mu \delta_{\rm e} f (q,p)~ dq dp = 0. 
\end{equation}
In particular, for the HMF model, 
the spatially inhomogeneous solution $f_0(q,p)$ is 
canonically formally stable if and only if
\end{claim}
\begin{equation}
	\label{eq:CFS}
	\begin{split}
		\mathcal{I}_{\rm C}[f_0] \equiv&
		1 + \int_{-\pi}^{\pi} dq \cos^2 q
		\int_{-\infty}^{\infty} \frac{1}{p} \frac{\partial f_0}{\partial p}(q,p)~ dp 
		\\
		&
		- \frac{\displaystyle \left(\int_{-\pi}^{\pi} dq \cos q 
		\int_{-\infty}^{\infty} \frac{1}{p} \frac{\partial f_0}{\partial p}(q,p)~ dp \right)^2}
		{\displaystyle \int_{-\pi}^{\pi} dq  
		\int_{-\infty}^{\infty} \frac{1}{p} \frac{\partial f_0}{\partial p}(q,p)~ dp} 
		> 0. 
	\end{split}
\end{equation}
Satisfying the inequality \eqref{eq:CFS} is sufficient but not necessary for the formal stability. 
We will show the existence of stationary solutions $f_0$ which are not canonically formally stable, but 
formally stable in the most refined sense.
\subsection{Example: Family of distributions having metastable states}
\begin{figure}[t b]
	    \begin{center}
	        \includegraphics[width=6cm]{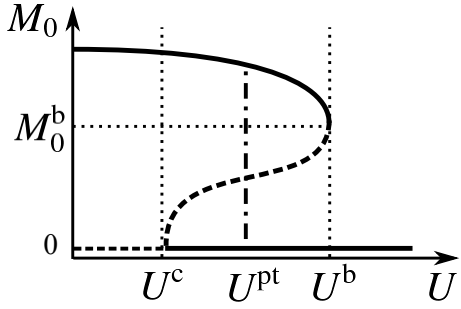}
	        \caption{
		Schematic picture of the phase diagram on $(M_0, U)$ for some fixed $M_{\rm I}$ \cite{AA07}. 
		The solid curve represents the stable or metastable states 
		which are realized as the local maximum points of the entropy. 
		The broken curve represents the unstable states which are realized as the local minimum points of the entropy. 
		These two curves meet at $(M_0^{\rm b}, U^{\rm b})$. 
		A region between $U^{\rm c}$ and $U^{\rm b}$ is called the two-phase coexistence region, and 
		the first order phase transition occurs at $U^{\rm pt}$. 
		}
	        \label{fig-coexist}
	    \end{center}
\end{figure}

In this subsection, we prove the following proposition: 
\begin{proposition}
\label{prop5}
Let $D$ be a subset of $\mathbb{R}^n$. 
Assume that a family of smooth stationary solutions 
$ \mathcal{X} = \{f_0(q, p; M_0, \lambda)~|~\lambda\in D \}$, 
which are parametrized with the order parameter $M_0 = M_x[f_0]$ and 
a set of macroscopic quantities $\lambda=(\lambda_1, \lambda_2, \cdots, \lambda_n) \in D$, 
such that there exists $f_0^{\rm b}(q,p) = f_0(q, p; M_0^{\rm b}, \lambda^{\rm b}) \in \mathcal{X}$ 
satisfying $\mathcal{I}\left[f_0^{\rm b}\right]= 0$
and $M\left[f_0^{\rm b}\right] > 0$. 
Moreover, assume that both $\mathcal{I}[f_0](M_0, \lambda)$ 
and $\mathcal{I}_{\rm C}[f_0](M_0, \lambda)$
depend on $M_0$ and $\lambda$ continuously. 
Then, there are stationary solutions $f_0 \in \mathcal{X}$ which do not satisfy 
the canonical formal stability criterion, $\mathcal{I}_{\rm C}[f_0](M_0, \lambda)> 0$ \eqref{eq:CFS}, 
but do satisfy the formal stability criterion,  
$\mathcal{I}[f_0](M_0, \lambda) =D_x(0) > 0$ \eqref{eq:f-stab}. 
\end{proposition}
\begin{remark}
If the system has a first order phase transition and a two-phase coexistence region 
in a parameter space $(M_0, \lambda)$, 
then we can take a family $\mathcal{X}$ of stationary solutions satisfying 
assumptions in Prop. \ref{prop5}. 
An example of such a family $\mathcal{X}$ is known in 
Lynden-Bell's distributions (or Fermi-Dirac type distributions)
\cite{DL67}. 
Within the Lynden-Bell's statistical mechanics 
with two-valued water-bag initial conditions, 
single-body distributions are parametrized with 
the order parameter in stationary states $M_0$, 
the energy $U$, and the parameter $M_{\rm I}$ describing to 
what extent particles spread on the $\mu$-space 
before violent relaxation occurs. 
In this case, one has $n = 2$ and 
$(\lambda_1, \lambda_2) = (U, M_\mathrm{I})$ 
\cite{AA07, FS11, SO11}. 
A schematic picture of the phase diagram $(M_0, U, M_{\rm I})$ is exhibited in Fig. \ref{fig-coexist}. 
On the three dimensional parameter space, 
one can observe a first order phase transition, 
a tricritical point, and a two-phase coexistence region. 
\end{remark}
We will omit the parameters $(M_0, \lambda)$ 
from the description of $f_0$, $\mathcal{I}[f_0]$ and $\mathcal{I}_{\rm C}[f_0]$ 
as long as no confusion arises. 
To prove the proposition, 
we first rewrite the third term of the right-hand side of $\eqref{eq:CFS}$ 
in terms of the angle-action coordinates, 
\begin{widetext}
\begin{equation}
	\frac{\displaystyle \left(\int_{-\pi}^{\pi} dq \cos q
	\int_{-\infty}^{\infty} \frac{1}{p} \frac{\partial f_0}{\partial p}(q,p)~ dp \right)^2}
	{\displaystyle \int_{-\pi}^{\pi} dq  
	\int_{-\infty}^{\infty} \frac{1}{p} \frac{\partial f_0}{\partial p}(q,p)~ dp}
	= 
	2 \pi \frac{\displaystyle \left( 
	\int_L \frac{\tilde{f}_0'(J)}{\Omega(J)} C^0(J)~ dJ \right)^2}
	{\displaystyle 
	\int_L \frac{\tilde{f}_0'(J)}{\Omega(J)}~ dJ}. 
\end{equation}
The difference between $\mathcal{I}[f_0]$ and $\mathcal{I}_{\rm C}[f_0]$ is calculated as
\begin{equation}
	\label{eq:differ-I-Ic}
	\begin{split}
		\mathcal{I}[f_0] - \mathcal{I}_{\rm C}[f_0] 
		&= 
		- 2 \pi \int_L \frac{\tilde{f}_0'(J)}{\Omega(J)} C^0(J)^2~dJ
		+ 2\pi \frac{\displaystyle \left( 
		\int_L \frac{\tilde{f}_0'(J)}{\Omega(J)} C^0(J)~ dJ \right)^2}
		{\displaystyle 
		\int_L \frac{\tilde{f}_0'(J)}{\Omega(J)}~ dJ} \\
		&= -2 \pi\int_L \frac{\tilde{f}_0'(J')}{\Omega(J')}~dJ'
		\left[\int_L C^0(J)^2 P(J)~dJ - 
		\left(\int_L C^0(J)P(J)~dJ\right)^2 \right], 
	\end{split}	
\end{equation}
\end{widetext}
where $P(J)$ is defined to be  
\begin{equation}
	\label{eq:prob}
	P(J) \equiv 
	\frac{1}{\displaystyle \int_L \frac{\tilde{f}_0'(J')}{\Omega(J')}~dJ'}
	\frac{\tilde{f}_0'(J)}{\Omega(J)} \geq 0.
\end{equation}
We note that the inequality, 
\begin{equation}
	\label{eq:verC}
	\int_L C^0(J)^2 P(J) ~dJ - 
	\left(\int_L C^0(J)P(J) ~dJ
	\right)^2 \geq 0,  
\end{equation}
is satisfied for any $f_0$. 
In fact, on account of 
\begin{equation}
	\int_L P(J)~dJ = 1,
\end{equation} 
we obtain the equation, 
\begin{equation}
	\begin{split}
		\label{eq:eqC}
		&\int_L C^0(J)^2 P(J) ~dJ - 
		\left(\int_L C^0(J)P(J) ~dJ
		\right)^2 \\
		&=\int_L 
		\left[
		C^0(J) - \int_L C^0(J') P(J')~dJ' 
		\right]^2
		P(J)~dJ, 
	\end{split}
\end{equation}
which implies Eq. \eqref{eq:verC}. 
If the equality holds in \eqref{eq:verC},
Eq. \eqref{eq:eqC} results in
\begin{equation}
	\label{eq:smooth-necessary}
	C^0(J) = \int_L C^0(J') P(J') ~dJ' = {\rm Constant}, \quad \forall J.
\end{equation}
However, this equality cannot be realized for any smooth spatially inhomogeneous solution, 
so that Eq. \eqref{eq:verC} should be
\begin{equation}
	\label{eq:verCneq}
	\int_L C^0(J)^2 P(J) ~dJ - 
	\left(\int_L C^0(J)P(J) ~dJ
	\right)^2 > 0.   
\end{equation}
Equation \eqref{eq:differ-I-Ic} 
with $\gamma(J) < 0$ and this inequality are 
put together to provide
\begin{equation}
	\label{eq:neq0}
	\mathcal{I}[f_0] > \mathcal{I}_{\rm C}[f_0] 
\end{equation}
for any smooth spatially inhomogeneous stationary solution $f_0$. 
This implies the known inclusion relation \cite{AC10}
\begin{equation}
	\begin{split}
		\{\textit{canonically form}&\textit{ally stable states}\}\\
		\cap&\\ 
		\{\textit{formally st}&\textit{able states}\}.
	\end{split}
\end{equation}

We show that there is a solution 
which is formally stable but not canonically formally stable. 
From the assumption in Prop. \ref{prop5}, 
\begin{equation}
	\label{eq:eqf}
	\mathcal{I} \left[f_0^{\rm b}\right] = 0. 
\end{equation}
If one could decide the formal stability 
of a stationary solution correctly by using the canonical formal stability criterion $\eqref{eq:CFS}$ 
near the stationary solution $f_0^{\rm b}$,  
the equation
\begin{equation}
	\label{eq:eq0}
	\mathcal{I}_{\rm C}\left[f_0^{\rm b}\right] = 0
\end{equation}
would be satisfied as well, 
since $\mathcal{I}_{\rm C}[f_0]$ depends on the parameters continuously. 
However, we have proved the inequality \eqref{eq:neq0}, 
so that \eqref{eq:eqf} and  \eqref{eq:eq0} do not hold simultaneously. 
Then the inequality 
\begin{equation}
	\label{eq:neq1}
	\mathcal{I}_{\rm C}\left[f_0^{\rm b}\right] < \mathcal{I}\left[f_0^{\rm b}\right] = 0
\end{equation}
should be satisfied. 
From \eqref{eq:neq0} and \eqref{eq:neq1}, 
it follows that there exists $f_0$ such that 
\begin{equation}
	\mathcal{I}_{\rm C} [f_0] \leq 0, \quad \mathcal{I}[f_0] > 0. 
\end{equation}
This implies that there is a solution which is formally stable, 
but not canonically formally stable. 

\section{Summary and Discussion}
\label{sec:conclusion}
We have worked out the spectral and formal stability criteria 
for spatially inhomogeneous stationary solutions to 
the Vlasov equation for the HMF model. 
These criteria are stated in the form of necessary and sufficient conditions 
(see Props. \ref{prop2} and \ref{prop-f-stab}). 
We stress that the assumptions for deriving the spectral stability criterion 
are satisfied by solutions to the variational equation $\eqref{eq:vareq}$.  
Our criterion avoids the problem of finding an infinite number of Lagrangian multipliers 
which are required in the previously obtained criterion \cite{AC10}. 
We note that the formal stability criterion in Prop. \ref{prop-f-stab} 
is stated in the form modified from the original one in \cite{DDH85}, 
since the perturbation $\delta_{\rm o} \tilde{f}$ perpendicular to the order parameter $\vec{M}_0$ 
with $M_y\left[\delta_{\rm o} \tilde{f}\right] \neq 0$ 
brings about the neutral formal stability, and since 
the set of neutrally formally stable solutions is defined so as not to be included in 
the set of formally stable ones by \cite{DDH85}. 

We have interpret the value of $\mathcal{I}[f_0] = D_x(0)$ 
as the strength of stability of the stable solutions.
Further, we have observed that 
the stationary state with high density almost harmonic orbits 
tends to be stable, 
and its stability gets to be stronger as $M_0$ gets large. 

We have shown that stability of some solutions in 
the family of stationary solutions having 
two-phase coexistence region in the phase diagram 
cannot be judged correctly by using 
the canonical formal stability criterion (see Prop. \ref{prop5}). 
A family of the Lynden-Bell's distributions is 
a family to which Prop. \ref{prop5} is applied.  

So far we have analyzed stability criteria for the HMF model without external fields. 
The present methods can be applied for the 
HMF model with non-zero external field, if the Hamiltonian takes the form
\begin{equation}
	\label{eq:HMF+h}
	\begin{split}
		&H_N = \sum_{i=1}^N \frac{p_i^2}{2} 
		+ \frac{1}{2N}\sum_{i,j=1}^N\left(1 - \cos(q_i - q_j)\right) - h \sum_{i = 1}^N \cos q_i.
	\end{split}
\end{equation}
All we have to do is to modify the single-body energy \eqref{e:1bodyE} by 
adding to the potential $-M_0 \cos q$ the term $-h \cos q$ coming from external field. 
Then, we can make a similar discussion by using the angle-action coordinates. 
In this case, the rotational symmetry is broken, so that 
$D_y(0) \neq 0$. 
Hence, the spectral and the formal stability criteria become  
\begin{equation}
	\label{eq:lin-nec}
	D_x(0) \geq 0, \quad D_y(0) \geq 0, 
\end{equation} 
and
\begin{equation}
	\label{eq:lin-suf}
	D_x(0) > 0, \quad D_y(0) > 0, 
\end{equation}
respectively, and further the value of $D_y(0)$ is computed as 
$D_y(0) = h/ (M_0+h)$ 
by using the same procedure as in \eqref{eq:Dy00}. 
In this case, 
the definition of formal stability is the same as one defined in \cite{DDH85}, 
so that we can refer to the linear stability condition. 
Equation. \eqref{eq:lin-nec} is a necessary condition 
for the linear stability of the spatially inhomogeneous solution, 
and \eqref{eq:lin-suf} is a sufficient condition of it. 
In fact, linearly stable states are spectrally stable states, 
and formally stable states are linearly stable states 
(see \cite{DDH85} for the proof). 
This discussion breaks down for 
the spatially inhomogeneous states in 
the HMF model without external field. 

The stability analysis performed in the present paper is applicable 
to the $\alpha$-HMF model $(0 \leq \alpha < 1)$ \cite{CA98} with the Hamitlonian
\begin{equation}
	\label{eq:alph-HMF+h}
	\begin{split}
		&H_N^\alpha = \sum_{i=1}^N \frac{p_i^2}{2} 
		- \kappa_N^\alpha \sum_{i>j}^N 
		\frac{ 1-\cos(q_i - q_j)}{|r_i -r_j|^\alpha} - h \sum_{i = 1}^N \cos q_i, 
	\end{split}
\end{equation}
where $r_i$ denotes the $i$-th lattice point, 
and the lattice spacing is set as $r_{i+1} - r_i = 1/N$. 
We assume the periodic boundary condition for the lattice, 
and the distance $|r_i - r_j|$ is actually $\min \left\{|r_i-r_j|, 1-|r_i - r_j| \right\}$. 
Then  $\kappa_N^\alpha$ is determined by
\begin{equation}
	\sum_{i = 1, i\neq j}^N \frac{\kappa_N^\alpha}{|r_i - r_j|^\alpha} = 1
\end{equation}
so that the system has the extensivity. 
Bachelard et al. \cite{RB11} have derived the Vlasov equation describing 
the dynamics of the $\alpha$-HMF model 
in the limit of infinite $N$. 
If the stationary state $f_0(q,p,r)$ does not depend on a configuration $r$ on the lattice, 
then the dispersion function can be written explicitly, and 
we can derive the spectral and formal stability criteria for the $\alpha$-HMF model 
as for the HMF model. 

Our procedure to look into the formal stability of the HMF model may be formally generalized 
to other models by using the biorthogonal functions and the Kalnajs' matrix form, 
which have been used in the astrophysics \cite{JBST, BOY11, AJK71}. 
However, there are difficulties in extending our result for the HMF model to 
that for general models. 
For instance, finding an appropriate biorthogonal system 
and analyzing the Kalnajs' matrix form are hard tasks. 
In fact, the dispersion function is not 
a complex-valued function but 
a linear operator or a matrix. 
Hence, the formal stability criterion 
should be described in the form of 
positive definiteness of matrices or linear operators. 
If this matrix is a diagonal matrix or a block diagonal matrix with small blocks, 
we may get the formal stability criterion as for the HMF model 
for each diagonal element or each block. 

\acknowledgments
The author is grateful to Professor Toshihiro Iwai and 
Professor Yoshiyuki Y. Yamaguchi 
for critical reading of this manuscript and 
for their helpful comments on this study. 
He is also grateful to Dr. Tarcisio N. Teles 
for his interest in this study and 
discussions. 
The author acknowledges the support of Grants for Excellent Graduate Schools,  
the hospitality of University of Florence, 
and the support of a JSPS Research Fellowship for Young Scientists. 
He also appreciates comments of 
an anonymous referee to improve the present paper.

\begin{appendix}

\section{Definition of the neutral stability}
\label{sec:neutral}
The neutral spectral stability is defined in terms of 
eigenvalues with vanishing real parts in Sec. \ref{sec:stability}. 
However, the neutral spectral stability is originally defined in terms of 
spectra, not eigenvalues only, with vanishing real parts \cite{DDH85}. 
The reason why we modify the definition of neutral stability 
is that the linear operator $\hat{\mathcal{L}}$ in \eqref{eq:lin} 
has always continuous spectrum on the imaginary axis, 
and this does not bring about spectral instability.

\section{Eigenvalues of the linearized Vlasov operator and roots of the dispersion relation}
\label{sec:app1}
We review the relation between eigenvalues of the linearized Vlasov operator 
\eqref{eq:lin} and roots of the dispersion relation after \cite{JDC89, NGVK55, KMC59}. 
We do not deal with continuous spectra, embedded eigenvalues, or Landau poles in this appendix, 
since they do not set off the spectral instability. 

\subsection{Spatially homogeneous state case}
Let $f_0(p)$ be a spatially homogeneous, even, unimodal, and smooth function. 
Let $\hat{\mathcal{L}}$ be the associated 
linearized Vlasov operator defined by \eqref{eq:lin}. 
Then, the linearized Vlasov equation around $f_0(p)$ takes the form 
\begin{equation}
	\label{eq:lin-Vlasov2}
	\frac{\partial f_1}{\partial t} = \hat{\mathcal{L}}[f_1] .
\end{equation}
We expand the both sides of \eqref{eq:lin-Vlasov2} into the Fourier series to find that 
the amplitude of the $k$-th Fourier mode $\hat{f}_{1,k}(p,t)$ obeys one of the following equations 
\begin{equation}
	\frac{\partial \hat{f}_{1,k}}{\partial t} =
	\begin{cases}
		0, &~ k = 0,\\
		\begin{split}
	\displaystyle
		&-ik\Bigg[ 
		p\hat{f}_{1,k}(p,t) \\
		&\qquad + \pi \int_{-\infty}^{\infty} \hat{f}_{1,k}(p',t)~dp' f_0'(p) 
		\Bigg], 
		\end{split}
		&~|k| = 1,\\
		-ikp \hat{f}_{1,k}(p,t), &~|k| \geq 2.
	\end{cases}
\end{equation}
For $|k| \geq 2$, 
there is no growth or damping mode. 
For $|k| = 1$, 
if $\lambda$ is an eigenvalue of the linearized Vlasov operator $\hat{\mathcal{L}}$, 
the associated eigenfunction can be written as $\tilde{f}_{1,k}(p) e^{\lambda t}$, and 
we get the equation for $\tilde{f}_{1,k}(p)$
\begin{equation}
	\label{eq:dD1}
	\tilde{f}_{1,k}(p) 
	= -\frac{\pi f_0'(p)}{p-i \lambda/k} 
	\int_{-\infty}^{\infty} \tilde{f}_{1,k}(p')~dp'. 
\end{equation} 
Integrating this equation over the whole $\mathbb{R}$ results in
\begin{equation}
	\label{eq:dD2}
	\int_{-\infty}^{\infty} \tilde{f}_{1,k}(p')~dp' 
	\left[1 + \pi \int_{-\infty}^{\infty} \frac{f_0'(p)}{p-i \lambda/k} dp
	\right] = 0,  
\end{equation}
which is rewritten as 
\begin{equation}
	\label{eq:dD3}
	\Lambda(i\lambda/k) \int_{-\infty}^{\infty} \tilde{f}_{1,k}(p')~dp' =0,
\end{equation}
where $\Lambda$ is defined to be 
\begin{equation}
	\Lambda(\omega) =1 + \pi \int_{-\infty}^{\infty} \frac{f_0'(p)}{p-\omega} dp,
\end{equation}
and is called the spectral function defined on $\mathbb{C}\setminus \mathbb{R}$ \cite{JDC89}. 
In view of \eqref{eq:disp-hom}, we find 
that $\Lambda(\omega) = D(\omega)$ on the upper half $\omega$-plane. 
Here we note that the relation between $\Lambda(\omega)$ and $D(\omega)$ for 
$\omega \in \mathbb{C} \setminus \mathbb{R}$ \cite{JDC89} is given by
\begin{equation}
	D(\omega) = 
	\begin{cases}
		\Lambda(\omega), &~ {\rm Im} \omega > 0, \\
		\Lambda(\omega)+ 2i\pi f_0'(\omega), &~{\rm Im} \omega < 0.
	\end{cases}
\end{equation}
If the factor including $\tilde{f}_{1,k}$ in \eqref{eq:dD2} vanishes, i.e., if 
\begin{equation}
	\label{eq:hatfk}
	\int_{-\infty}^{\infty} \tilde{f}_{1,k}(p')~dp'  = 0, 
\end{equation}
then $\tilde{f}_{1,k}(p)$ vanishes owing to $\eqref{eq:dD1}$, 
and thereby it has no concern with stability. 
We are then allowed to assume that the left-hand side 
of \eqref{eq:hatfk} do not vanish. 
It then follows from \eqref{eq:dD3} that 
if $\lambda$ is an eigenvalue of the linearized Vlasov operator $\hat{\mathcal{L}}$, 
the equation $\Lambda(i\lambda/k) = 0$ should be satisfied for $k =1$ or for $k = -1$. 
It is to be remarked that 
the assumptions imposed on $f_0$ in Prop. \ref{prop1} give rise to the relation, 
\begin{equation}
	\label{eq:Lambda-Lambda}
	\Lambda(\omega) = \Lambda(-\omega) = \Lambda(\omega^\ast)^\ast = 
	\Lambda(-\omega^\ast)^\ast. 
\end{equation}
This implies that if $\omega$ with ${\rm Im} \omega > 0$ is a root of the dispersion relation \eqref{eq:disp-hom}, 
then the linearized Vlasov operator $\hat{\mathcal{L}}$ has eigenvalues 
$i \omega$, $-i \omega$, $-i \omega^\ast$, and $i \omega^\ast$. 
Therefore, if $\hat{\mathcal{L}}$ has an eigenvalue, 
$\hat{\mathcal{L}}$ has inevitably an unstable eigenvalue, 
so that the solution $f_0$ should be unstable. 

\subsection{Spatially inhomogeneous state case}
So far we have analyzed the stability of spatially homogenous states. 
The procedure can be applied to spatially inhomogeneous states \cite{BOY10}. 
Let us rewrite the Poisson bracket as 
\begin{equation}
	\{a, b \} = 
	\frac{\partial a}{\partial J} \frac{\partial b}{\partial \theta}-
	\frac{\partial a}{\partial \theta}\frac{\partial b}{\partial J}  
\end{equation}
in terms of the angle-action coordinates. 
The linearized Vlasov equation can be written also 
in terms of the angle-action coordinates as follows,  
\begin{equation}
	\label{eq:lin-V-AA}
	\frac{\partial f_1}{\partial t} + \Omega(J) \frac{\partial f_1}{\partial \theta} 
	- f_0'(J) \frac{\partial}{\partial \theta} \mathcal{V}[f_1] = 0,  
\end{equation}
where $\Omega(J) = d \mathcal{E}(J)/dJ$. 
We omit to put the tilde over $f_0$ and $f_1$ to specify that 
the arguments of these functions are the angle-action variables, 
in this section.  
We expand the functions $f_1(\theta, J, t)$, $\cos q(\theta, J)$, and $\sin q(\theta, J)$ 
into the Fourier series,  
\begin{align}
	f_1(\theta, J, t) &=\sum_{k\in \mathbb{Z}} \hat{f}_{1,k}(J,t) e^{ik \theta}, \label{eq:f1-Fourier}\\
	\cos q(\theta, J)&=\sum_{k\in \mathbb{Z}} C^k(J) e^{ik \theta},\label{eq:cos-Fourier}\\
	\sin q(\theta, J)&=\sum_{k\in \mathbb{Z}} S^k(J) e^{ik \theta},\label{eq:sin-Fourier}
\end{align}
respectively. 
By using \eqref{eq:f1-Fourier}, \eqref{eq:cos-Fourier}, and \eqref{eq:sin-Fourier}, 
the potential term $\mathcal{V}[f_1]$ in \eqref{eq:lin-V-AA} is rewritten as
\begin{equation}
	\label{eq:potential-f1}
	\begin{split}
		&-\mathcal{V}[f_1] \left(q(\theta, J)\right)\\
		&= 
		\iint_\mu \cos\left(q(\theta,J)- q'(\theta',J') \right) f_1(\theta', J', t)~ d\theta' dJ'\\
		&= 2\pi
		\sum_{m \in \mathbb{Z}} C^m(J) e^{im \theta} \sum_{k \in \mathbb{Z}} 
		\int_L {C^k}(J')^\ast \hat{f}_{1,k}(J',t)~dJ' \\
		&\quad+
		2\pi
		\sum_{m \in \mathbb{Z}} S^m(J) e^{im \theta} \sum_{k \in \mathbb{Z}} 
		\int_L {S^k}(J')^\ast \hat{f}_{1,k}(J',t)~dJ'.
	\end{split}
\end{equation}
Then, the $m$-th Fourier mode $\hat{f}_{1,m}$ is shown to satisfy the equation, 
\begin{equation}
	\label{eq:lin-V-AAm}
	\begin{split}
		&\frac{\partial \hat{f}_{1,m}}{\partial t} 
		=
		- i m \Omega(J) \hat{f}_{1,m}(J, t) \\
		&- 2 \pi im C^m(J) f_0'(J) 
		 \sum_{k \in \mathbb{Z}} 
		\int_L {C^k}(J')^\ast \hat{f}_{1,k}(J',t)~dJ' \\
		&- 2\pi im S^m(J) f_0'(J)
		 \sum_{k \in \mathbb{Z}} 
		\int_L {S^k}(J')^\ast \hat{f}_{1,k}(J',t)~dJ'.
	\end{split}
\end{equation}
Let $\hat {f}_{1}^\lambda(\theta, J,t) = \sum_{m \in \mathbb{Z}} \hat{f}_{1,m}^\lambda(J,t)e^{im \theta}$ 
be an eigenfunction associated with an eigenvalue $\lambda$ of $\hat{\mathcal{L}}$, i.e., 
$\left(\hat{f}_{1,m}^\lambda \right)_{m \in \mathbb{Z}}$ be the eigenvector 
associated with the eigenvalue $\lambda$. 
Setting
\begin{equation}
	\label{eq:eigen-vec1}
	\hat{f}_{1,m}^\lambda (J, t) = \tilde{f}_{1,m}^\lambda (J) e^{\lambda t}, ~\forall m \in \mathbb{Z}, 
\end{equation}
and substituting it into $\eqref{eq:lin-V-AAm}$, 
we get 
\begin{equation}
	\label{eq:lin-V-AA-eigen}
	\begin{split}
	\tilde{f}_{1,m}^\lambda (J) =&
	- 2\pi \frac{m f_0'(J)}{m \Omega(J) - i \lambda} 
	C^m(J) \sum_{k \in \mathbb{Z}}
	\int_L C^k(J')^\ast \tilde{f}_{1,k}^\lambda(J')~dJ'\\
	&- 2\pi \frac{m f_0'(J)}{m \Omega(J) - i \lambda} 
	S^m(J) \sum_{k \in \mathbb{Z}}
	\int_L S^k(J')^\ast \tilde{f}_{1,k}^\lambda(J')~dJ'.
	\end{split}
\end{equation}
Multiplying $C^m(J)^\ast$ or $S^m(J)^\ast$ to both sides of \eqref{eq:lin-V-AA-eigen}, 
summing up over $m \in \mathbb{Z}$, 
and using the fact \cite{BOY10}
\begin{equation}
	2 \pi \sum_{m \in \mathbb{Z}} \int_L 
	 \frac{m f_0'(J)}{m \Omega(J) - i \lambda}C^m(J)^\ast S^m(J) ~dJ = 0, 
\end{equation}
we obtain the equations 
\begin{equation}
	\begin{split}
		\left[1 + 2 \pi \sum_{m\in \mathbb{Z}} \int_L   \frac{m f_0'(J)}{m \Omega(J) - i \lambda} 
		\left|C^m(J)\right|^2~dJ\right] &\\
		\times \sum_{k \in \mathbb{Z}} 
		\int_L C^k(J')^\ast \tilde{f}_{1,k}^\lambda(J')~dJ' 
		&= 0,
	\end{split}
\end{equation}
and 
\begin{equation}
	\begin{split}
		\left[1 + 2 \pi \sum_{m\in \mathbb{Z}} \int_L   \frac{m f_0'(J)}{m \Omega(J) - i \lambda} 
		\left|S^m(J)\right|^2~dJ\right]& \\
		\times \sum_{k \in \mathbb{Z}} 
		\int_L S^k(J')^\ast \tilde{f}_{1,k}^\lambda(J')~dJ' 
		&= 0.
	\end{split}
\end{equation}
A necessary condition for the existence of 
the non-zero eigenfunction corresponding to the eigenvalue $\lambda$ is 
that at least one of the following two equations is satisfied;
\begin{equation}
	\label{eq:Lx-Ly}
	\begin{split}
	\Lambda_x(i\lambda) &\equiv 
	1 + 2 \pi \sum_{m\in \mathbb{Z}} \int_L   \frac{m f_0'(J)}{m \Omega(J) - i \lambda} 
	\left|C^m(J)\right|^2~dJ
	= 0, \\
	\Lambda_y(i\lambda) &\equiv
	1 + 2 \pi \sum_{m\in \mathbb{Z}} \int_L   \frac{m f_0'(J)}{m \Omega(J) - i \lambda} 
	\left|S^m(J)\right|^2~dJ 
	= 0.
	\end{split} 
\end{equation}
When ${\rm Im \omega} > 0$,
the spectral functions $\Lambda_x(\omega)$ and $\Lambda_y(\omega)$ 
defined in \eqref{eq:Lx-Ly} coincide with the dispersion functions 
$D_x(\omega)$ and $D_y(\omega)$ 
defined in \eqref{eq:inhom-disp} and \eqref{eq:inhom-disp-perp}, 
respectively. 
As in the homogeneous state case, 
both $\Lambda_x(\omega)$ and $\Lambda_y(\omega)$
satisfy the relation \eqref{eq:Lambda-Lambda}, 
since $|C^m(J)| = |C^{-m}(J)|$ and $|S^m(J)| = |S^{-m}(J)|$ 
are satisfied for all $m \in \mathbb{Z}$. 
It turns out that if $\omega$ with ${\rm Im} \omega > 0$ is 
a root of the dispersion relation $D_x(\omega) = 0$ or $D_y(\omega) = 0$, 
the linearized Vlasov operator has eigenvalues, 
$i \omega$, $-i \omega$, $-i \omega^\ast$, and $i \omega^\ast$. 
We hence conclude that if $\hat{\mathcal{L}}$ has an eigenvalue, 
$\hat{\mathcal{L}}$ has inevitably an unstable eigenvalue, 
so that the stationary solution $f_0(J)$ should be unstable.

\end{appendix}

\end{document}